%% file: main.tex
\pdfoutput=1

\documentclass[12pt,a4paper]{article}

\usepackage{ifthen} 
\newboolean{pdflatex}
\setboolean{pdflatex}{true} 
\newboolean{articletitles}
\setboolean{articletitles}{true} 
\newboolean{uprightparticles}
\setboolean{uprightparticles}{false} 
\newboolean{inbibliography}
\setboolean{inbibliography}{false} 
\usepackage{microtype}
\usepackage{lineno}  
\usepackage{xspace} 
\usepackage{caption}

\usepackage{graphicx}  
\usepackage{color}
\usepackage{colortbl}
\usepackage{amsmath} 
\usepackage{amssymb}
\usepackage{amsfonts}
\usepackage{upgreek} 
\newcommand*\patchAmsMathEnvironmentForLineno[1]{%
\expandafter\let\csname old#1\expandafter\endcsname\csname #1\endcsname
\expandafter\let\csname oldend#1\expandafter\endcsname\csname
end#1\endcsname
 \renewenvironment{#1}%
   {\linenomath\csname old#1\endcsname}%
   {\csname oldend#1\endcsname\endlinenomath}%
}
\newcommand*\patchBothAmsMathEnvironmentsForLineno[1]{%
  \patchAmsMathEnvironmentForLineno{#1}%
  \patchAmsMathEnvironmentForLineno{#1*}%
}
\AtBeginDocument{%
\patchBothAmsMathEnvironmentsForLineno{equation}%
\patchBothAmsMathEnvironmentsForLineno{align}%
\patchBothAmsMathEnvironmentsForLineno{flalign}%
\patchBothAmsMathEnvironmentsForLineno{alignat}%
\patchBothAmsMathEnvironmentsForLineno{gather}%
\patchBothAmsMathEnvironmentsForLineno{multline}%
\patchBothAmsMathEnvironmentsForLineno{eqnarray}%
}
\usepackage{hyperref}    
\usepackage[all]{hypcap} 
\input{lhcb-symbols-def}

\usepackage{cite} 
\usepackage{mciteplus}
\usepackage{longtable} 
\usepackage{subfigure}
\usepackage{dcolumn}
\usepackage{multirow}
\usepackage{booktabs}
\usepackage[percent]{overpic}

\input{bhhh-symbols-def}       
\def\kpipisigma  {\ensuremath{2.8}\xspace}
\def\kkksigma    {\ensuremath{4.3}\xspace}
\def\pipipisigma {\ensuremath{4.2}\xspace}
\def\kkpisigma   {\ensuremath{5.6}\xspace}
\def\kpipiacp    {\ensuremath{0.025 \pm 0.004 \pm 0.004 \pm 0.007}\xspace}                
\def\kkkacp      {\ensuremath{-0.036 \pm 0.004 \pm 0.002 \pm 0.007}\xspace}              
\def\pipipiacp   {\ensuremath{0.058 \pm 0.008 \pm 0.009 \pm 0.007}\xspace}            
\def\kkpiacp     {\ensuremath{-0.123 \pm 0.017 \pm 0.012 \pm 0.007}\xspace}

\begin{document}

\renewcommand{\thefootnote}{\fnsymbol{footnote}}
\setcounter{footnote}{1}

\begin{titlepage}
\pagenumbering{roman}

\vspace*{-1.5cm}
\centerline{\large EUROPEAN ORGANIZATION FOR NUCLEAR RESEARCH (CERN)}
\vspace*{1.5cm}
\hspace*{-0.5cm}
\begin{tabular*}{\linewidth}{lc@{\extracolsep{\fill}}r}
\ifthenelse{\boolean{pdflatex}}
{\vspace*{-2.7cm}\mbox{\!\!\!\includegraphics[width=.14\textwidth]{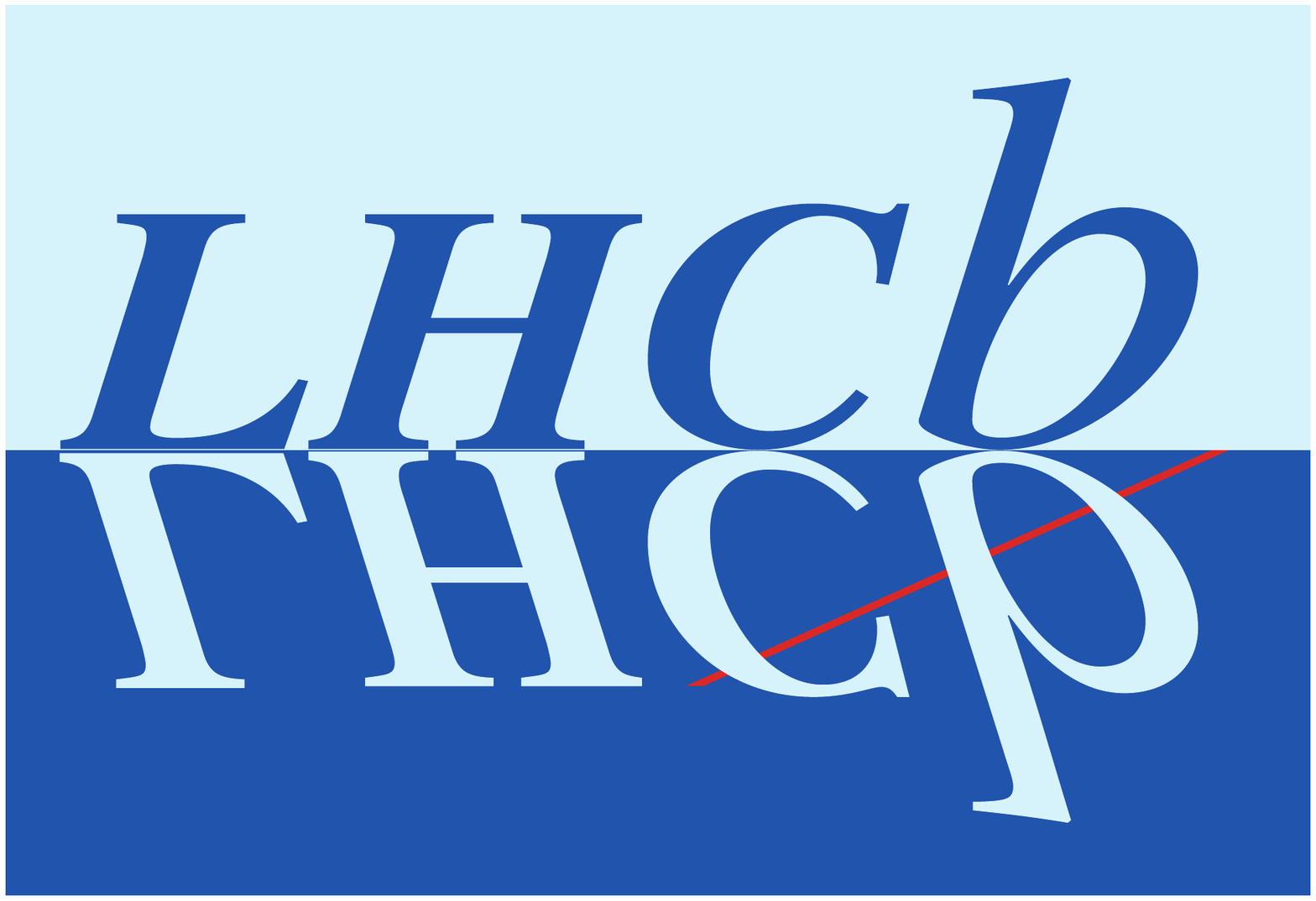}} & &}%
{\vspace*{-1.2cm}\mbox{\!\!\!\includegraphics[width=.12\textwidth]{lhcb-logo.eps}} & &}%
\\
 & & CERN-PH-EP-2014-203 \\  
 & & LHCb-PAPER-2014-044 \\  
 & & December 12, 2014 \\
 & & \\
\end{tabular*}

\vspace*{2.0cm}

{\bf\boldmath\huge
\begin{center}
Measurements of $C\!P$ violation \\in the three-body  phase space of charmless $B^{\pm}$ decays
\end{center}
}

\vspace*{1.5cm}

\begin{center}
The LHCb collaboration\footnote{Authors are listed at the end of this paper.}
\end{center}

\vspace*{.5cm}

\begin{abstract}
  \noindent
The charmless three-body decay modes \ensuremath{{B^{\pm} \rightarrow K^{\pm} \pi^{+} \pi^{-}}}, \ensuremath{{B^{\pm} \rightarrow K^{\pm} K^{+} K^{-}}}, \ensuremath{{B^{\pm} \rightarrow \pi^{\pm} K^{+} K^{-}}} and \ensuremath{{B^{\pm} \rightarrow \pi^{\pm} \pi^{+} \pi^{-}}} are reconstructed using data, corresponding to an integrated luminosity of 3.0\;\ensuremath{\mbox{\,fb}^{-1}}, collected by the LHCb detector.
The inclusive \ensuremath{C\!P} asymmetries of these modes are measured to be
\begin{eqnarray}
\acp(\ensuremath{{B^{\pm} \rightarrow K^{\pm} \pi^{+} \pi^{-}}})&=& +\ensuremath{0.025 \pm 0.004 \pm 0.004 \pm 0.007} \, , \nonumber \\
\acp(\ensuremath{{B^{\pm} \rightarrow K^{\pm} K^{+} K^{-}}}) &=& \ensuremath{-0.036 \pm 0.004 \pm 0.002 \pm 0.007} \, , \nonumber \\
\acp(\ensuremath{{B^{\pm} \rightarrow \pi^{\pm} \pi^{+} \pi^{-}}})&=& +\ensuremath{0.058 \pm 0.008 \pm 0.009 \pm 0.007} \, , \nonumber \\
\acp(\ensuremath{{B^{\pm} \rightarrow \pi^{\pm} K^{+} K^{-}}})&=& \ensuremath{-0.123 \pm 0.017 \pm 0.012 \pm 0.007} \, , \nonumber
\end{eqnarray}
where the first uncertainty is statistical, the second systematic, and the third is due to the \ensuremath{C\!P} asymmetry of the \ensuremath{{B^{\pm} \rightarrow \jpsi K^{\pm}}} reference mode.
The distributions of these asymmetries are also studied as functions of position in the Dalitz plot and suggest contributions from rescattering and resonance interference processes.
\end{abstract}

\vspace*{1.0cm}

\begin{center}
  Submitted to Phys.~Rev.~D
\end{center}

\vspace{\fill}

{\footnotesize
\centerline{\copyright~CERN on behalf of the \lhcb collaboration, license \href{http://creativecommons.org/licenses/by/4.0/}{CC-BY-4.0}.}}
\vspace*{2mm}

\end{titlepage}

\newpage
\setcounter{page}{2}
\mbox{~}

\cleardoublepage

\renewcommand{\thefootnote}{\arabic{footnote}}
\setcounter{footnote}{0}

\pagestyle{plain} 
\setcounter{page}{1}
\pagenumbering{arabic}

\section{Introduction}

The violation of  \CP symmetry is well established experimentally in the quark sector and,
in the Standard Model (SM), is explained by the Cabibbo-Kobayashi-Maskawa~\cite{CKM} matrix through the presence of a single irreducible complex phase.
Although the SM is able to describe all \CP asymmetries observed experimentally in particle decays, the amount of \CP violation within the SM is insufficient to explain the matter-antimatter asymmetry of the universe~\cite{Gavela:1993ts}.

The decays of $B$ mesons with three charged charmless mesons in the final state offer interesting opportunities to search for different sources of \CP violation, through the study of the signature of these sources in the Dalitz plot.
Several theoretical studies modelled the dynamics of the decays in terms of two-body intermediate states, such as $\rho(770)\Kpm$ or $\KorKbar^{*0}(892)\pipm$ for \kpipi decays, and $\phi(1020) \Kpm$ for \kkk decays (see \eg~\cite{Neubert}).
These intermediate states were identified through amplitude analyses in which a resonant model was assumed.
One method of performing such analyses was used by the Belle and the BaBar collaborations and significant \CP violation was  observed in the intermediate $\rho^0\Kpm$ state~\cite{bellek2pi,BaBark2pi} and in the $\phi \Kpm$ channel~\cite{BaBarkkk}.
No significant inclusive \CP asymmetry (integrated over the Dalitz plot) was found in \kpipi or \kkk decays~\cite{bellek2pi,BaBarkkk}.
Another method is to measure the \CP asymmetry in different regions of the three-body phase space.
The LHCb collaboration measured non-zero inclusive \CP asymmetries and larger local asymmetries in the decays
\kpipi, \kkk~\cite{LHCb-PAPER-2013-027}, \kkpi and \pipipi~\cite{LHCb-PAPER-2013-051} using a sample corresponding to 1.0\invfb of data.
These results suggested that final-state interactions may be a contributing factor to \CP violation~\cite{Bhattacharya:2013cvn,IgnacioCPT}.

Direct \CP violation requires the existence of amplitudes with differences in both their weak and their strong phases.
The value of the weak phase can be accessed through interference between tree-level contributions to charmless $B$ decays and other amplitudes ({\it e.g.} penguins).
The strong phase can originate from three different sources in charmless three-body decays.
The first source is related to short-distance processes where the gluon involved in the penguin contribution is timelike, \ie the momentum transfer satisfies $q^{2} > 4m^{2}_{i}$, where $m_i$ represents the mass of either the $u$ or the $c$ quark present in the loop diagram~\cite{Gerard}.
This process is similar to that proposed for two-body decays where  \CP violation is caused by short-distance processes~\cite{BSS1979}.
The remaining two sources are related to long-distance effects involving hadron-hadron interactions in the final state.
Interference between intermediate states of the decay can introduce large strong-phase differences, and therefore induce local asymmetries in the phase space~\cite{Miranda1, BGM,Bhattacharya:2013cvn,Lesniak2014201,PhysRevD87076007}.
Another mechanism is final-state $KK \leftrightarrow \pi \pi$ rescattering, which can occur between decay channels having the same flavour quantum numbers~\cite{LHCb-PAPER-2013-027,LHCb-PAPER-2013-051,Bhattacharya:2013cvn,IgnacioCPT}.
Conservation of \CPT symmetry constrains hadron rescattering so that the sum of the partial decay widths of all channels with the same final-state quantum numbers related by the scattering matrix must equal that of their charge-conjugated decays~\cite{PhysRevD.88.114014}.
The effects of SU(3) flavour symmetry breaking have also been investigated and can explain part of the pattern of \CP violation reported by LHCb~\cite{Xu:2013dta,Bhattacharya:2013cvn,Gronau:2013mda,PhysRevD.88.114014}.

In this paper, the inclusive \CP asymmetries of \kpipi, \kkk, \kkpi and \pipipi decays (henceforth collectively referred to as \hhh decays) are measured, and local asymmetries in specific regions of the phase space are studied.
All asymmetries are measured using the \jpsik channel, which has similar topology and negligible \CP violation, as a reference, thus allowing corrections to be made for production and instrumental asymmetries.
We use a sample of proton-proton collisions collected in 2011 (2012) at a centre-of-mass energy of 7(8) TeV and corresponding to an integrated luminosity of 1.0 (2.0)~\invfb.
This analysis supersedes that of~\cite{LHCb-PAPER-2013-027,LHCb-PAPER-2013-051}, by using a larger data sample, improved particle identification and a more performant event selection.

\section{LHCb detector and data set}
\label{dataset}

The \lhcb detector~\cite{Alves:2008zz} is a single-arm forward spectrometer covering the \mbox{pseudorapidity} range $2<\eta <5$, designed for the study of particles containing \bquark or \cquark
quarks.
The detector includes a high-precision tracking system consisting of a silicon-strip vertex detector surrounding the $pp$ interaction region, a large-area silicon-strip detector located upstream of a dipole magnet with a bending power of about $4{\rm\,Tm}$, and three stations of silicon-strip detectors and straw drift tubes placed downstream.
The tracking system provides a measurement of momentum, \ptot,  with a relative uncertainty that varies from 0.4\% at low momentum to 0.6\% at 100\gevc.
The minimum distance of a track to a primary vertex, the impact parameter, is measured with a resolution of $(15+29/\pt)\mu m$, where \pt is the component of \ptot transverse to the beam, in \gevc.
Charged hadrons are identified using two ring-imaging Cherenkov (RICH) detectors  \cite{LHCb-DP-2012-003}.
Photon, electron and hadron candidates are identified by a calorimeter system consisting of scintillating-pad and preshower detectors, an electromagnetic calorimeter and a hadronic calorimeter.
Muons are identified by a system composed of alternating layers of iron and multiwire proportional chambers~\cite{LHCb-DP-2012-002}.

The trigger~\cite{LHCb-DP-2012-004} consists of a hardware stage, based on information from the calorimeter and muon systems, followed by a software stage, which applies  full event reconstruction.
At the hardware trigger stage,  events are required to have a muon with  high \pt or a hadron, photon or electron with high transverse energy in the calorimeters.
For hadrons, the transverse energy threshold is 3.5\gev.
In this analysis two partially overlapping categories of events selected by the hardware trigger are considered: events where one of the hadrons from the \Bpm decay is used in the trigger decision (the ``trigger on signal'' sample), and events that are triggered by particles other than those hadrons from the \Bpm decay (the ``trigger independent of signal'' sample).

At the software trigger stage, events must have at least one good quality track from the signal decay candidate with high \pt and a significant displacement from any primary vertex (PV). 
A secondary vertex, consisting of three good quality tracks that have significant displacements from any PV, is also required.

The magnetic field polarity is reversed regularly during the data taking to reduce any potential bias from charged particle and antiparticle detection asymmetries.
The magnetic field bends charged particles in the horizontal plane and the two polarities are referred to as ``up" and ``down".
The fraction of data collected with the magnet down polarity is approximately 60\% in 2011, and 52\% in 2012.

Possible residual charge-dependent asymmetries, which may originate from left-right differences in detection efficiency, are studied by comparing measurements from data with inverted magnet polarities and found to be negligible.
Since the detection and production asymmetries are expected to change between 2011 and 2012 due to different data taking conditions, the analysis is carried out separately for the 2011 and 2012 data and the results combined.

The simulated events are generated using \pythia~8~\cite{Sjostrand:2006za} with a specific \lhcb configuration~\cite{LHCb-PROC-2010-056}.
Decays of hadronic particles are produced by \evtgen~\cite{Lange:2001uf}, in which final-state radiation is generated using \photos~\cite{Golonka:2005pn}.
The interaction of the generated particles with the detector and its response are implemented using the \geant toolkit~\cite{Allison:2006ve, *Agostinelli:2002hh} as described in Ref.~\cite{LHCb-PROC-2011-006}.

\section{Event selection}
Since the four \Bpm signal decay modes considered are topologically and kinematically similar, the same selection criteria are used for each, except for the particle identification requirements, which are specific to each final state.
The decay \jpsik, \jpsimumu serves as a control channel for \hhh decay modes.
Since it has negligible \CP violation, the raw asymmetry observed in  \jpsik decays is entirely due to production and detection asymmetries.
The control channel has a similar topology to the signal and the sample passes the same trigger, kinematic, and kaon particle identification selection as the signal samples.
The kaons from \jpsik decays also have similar kinematic properties in the laboratory frame to those from the \kpipi and \kkk modes.

In a preselection stage, loose requirements are imposed on the \ptot, \pt and the displacement from any PV for the tracks,
and  on the distance of closest approach between each pair of tracks.
The three tracks must form a good quality secondary vertex that has a significant separation from its associated PV.
The  momentum vector of the reconstructed $B^{\pm}$ candidate has to point back to the PV.

Charm meson contributions are removed by excluding events where two-body invariant masses \mpipi, \mkpi and \mkk are within  30\mevcc of the known value of the $\Dz$ mass~\cite{PDG2012}.
The contribution of misidentified \jpsik decays is also excluded from the \kpipi sample by removing the mass region $3.05 <\mpipi<3.15\gevcc$.

A multivariate selection based on a boosted decision tree (BDT) algorithm~\cite{Breiman,Roe,TMVA2007} is applied to reduce the combinatorial background.
The input variables, which are a subset of those used in the preselection, are common to all four decay modes.
The BDT is trained using a mixture of simulated signal events as the signal sample, and events reconstructed as \pipipi decays with  $5.40 < m(\pipm\pip\pim) < 5.58~\gev/c^2$ as the background sample.
The requirement on the BDT response is chosen to maximise the  ratio $N_S/\sqrt{N_S+N_B}$, where $N_S$ and $N_B$ represent the expected number of signal and background candidates, respectively, within an invariant mass window of approximately 40\mevcc around the signal peak.
Since the optimal requirements are similar for the different channels, the same BDT response requirement is chosen for all channels, to simplify the evaluation of the systematic uncertainties.
The BDT selection improves the efficiencies for selecting signal events by approximately 50\%, compared to the cut-based selection used in Refs.~\cite{LHCb-PAPER-2013-027} and~\cite{LHCb-PAPER-2013-051}.

Particle identification is used to reduce the cross-feed from other $B$ decays in which hadrons are incorrectly classified.
The main source is $K \to \pi$ and $\pi\to K$ misidentification, while $p\to K$ and $p\to \pi$ misidentification is negligible.
Muons are rejected by a veto applied to  each track~\cite{LHCb-DP-2013-001}.
After the full selection, events with more than one candidate  in the range $4.8 < m(B) < 5.8 \gevcc$ are discarded.
This removes approximately 1--2\% of candidates.

The \jpsik control channel is selected using the same criteria as described above, with two exceptions that enhance the selection of \jpsi mesons decaying to two muons: criteria used to identify charged pions are removed and the requirement  $3.05 < \mpipi < 3.15\gevcc$ is applied.

\section{Determination of signal yields}

For each channel the yields and raw asymmetry are extracted from a single simultaneous unbinned extended maximum likelihood fit to the \Bp and \Bm invariant mass distribution.
The signal components of all four channels are parametrised by a Gaussian function with widths and tails that differ either side of the peak to account for asymmetric effects such as final-state radiation.
The means and widths are allowed to vary in the fits, while the tail parameters are fixed to values obtained from simulation.      
The combinatorial backgrounds are described by exponential functions.
The backgrounds due to partially reconstructed four-body \B decays are parametrised by an ARGUS function~\cite{Argus} convolved with a Gaussian function.
The shapes and yields of peaking backgrounds, \ie fully reconstructed $B$ decays with at least one misidentified particle in the final state, are obtained from simulation  of the relevant decay modes and fixed in the fits.
The yields of the peaking and partially reconstructed background components are constrained to be equal for \Bp and \Bm decays.

\begin{figure}[tb]
\begin{overpic}[width=0.49\linewidth]{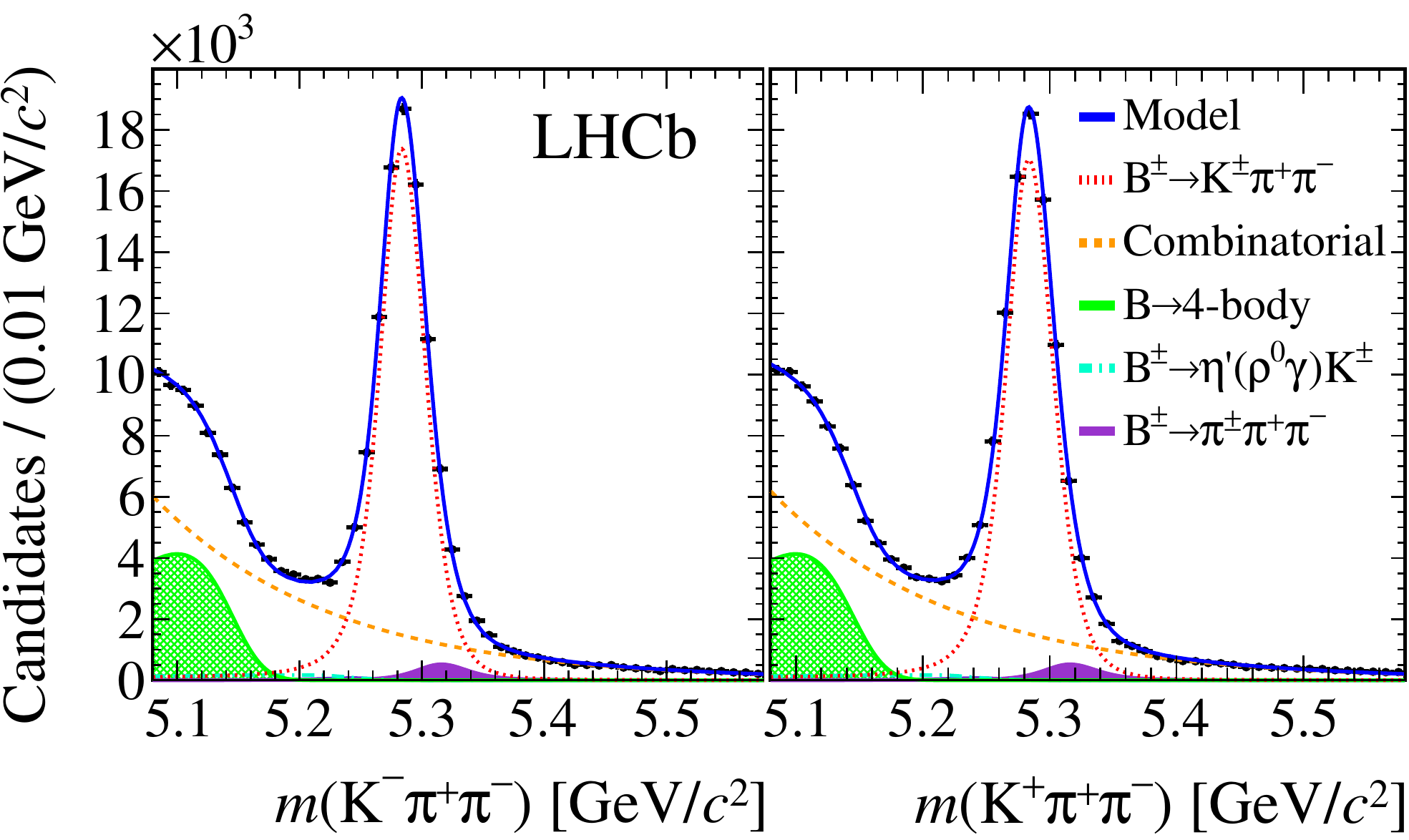}
 \put(15,45){\bf{(a)}}
\end{overpic}
\begin{overpic}[width=0.49\linewidth]{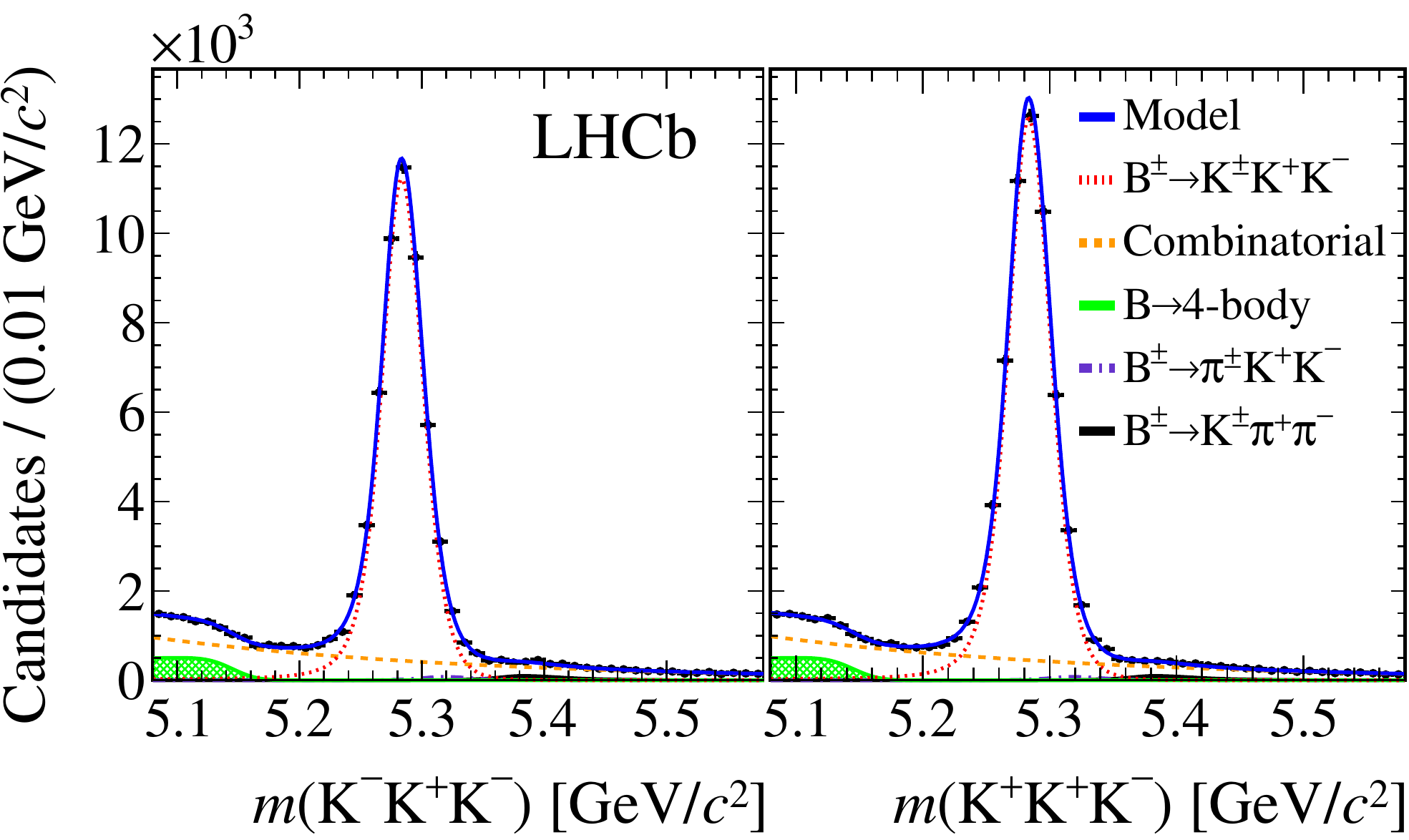}
 \put(15,45){\bf{(b)}}
\end{overpic}
\begin{overpic}[width=0.49\linewidth]{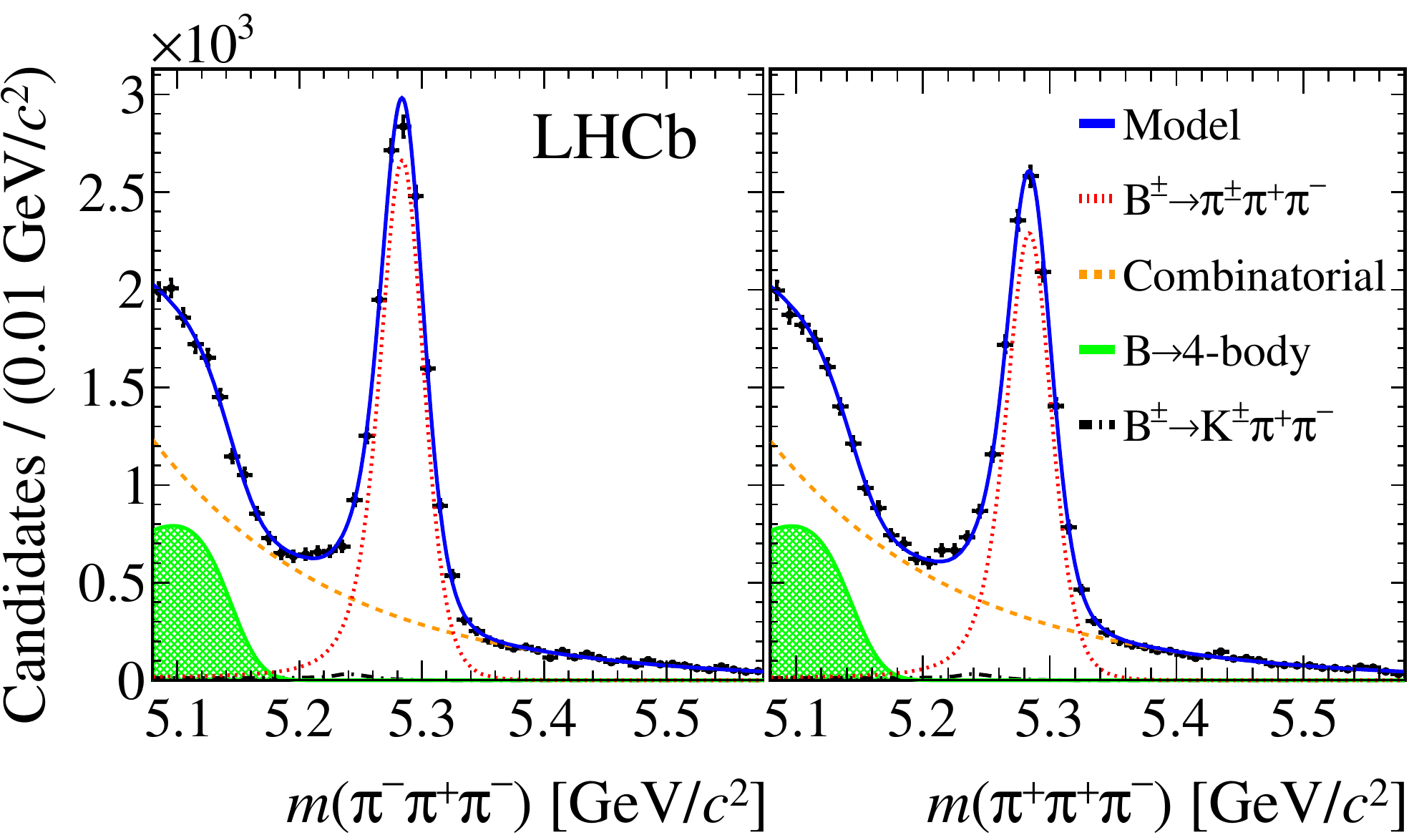}
 \put(15,45){\bf{(c)}}
\end{overpic}
$\;\;$\begin{overpic}[width=0.49\linewidth]{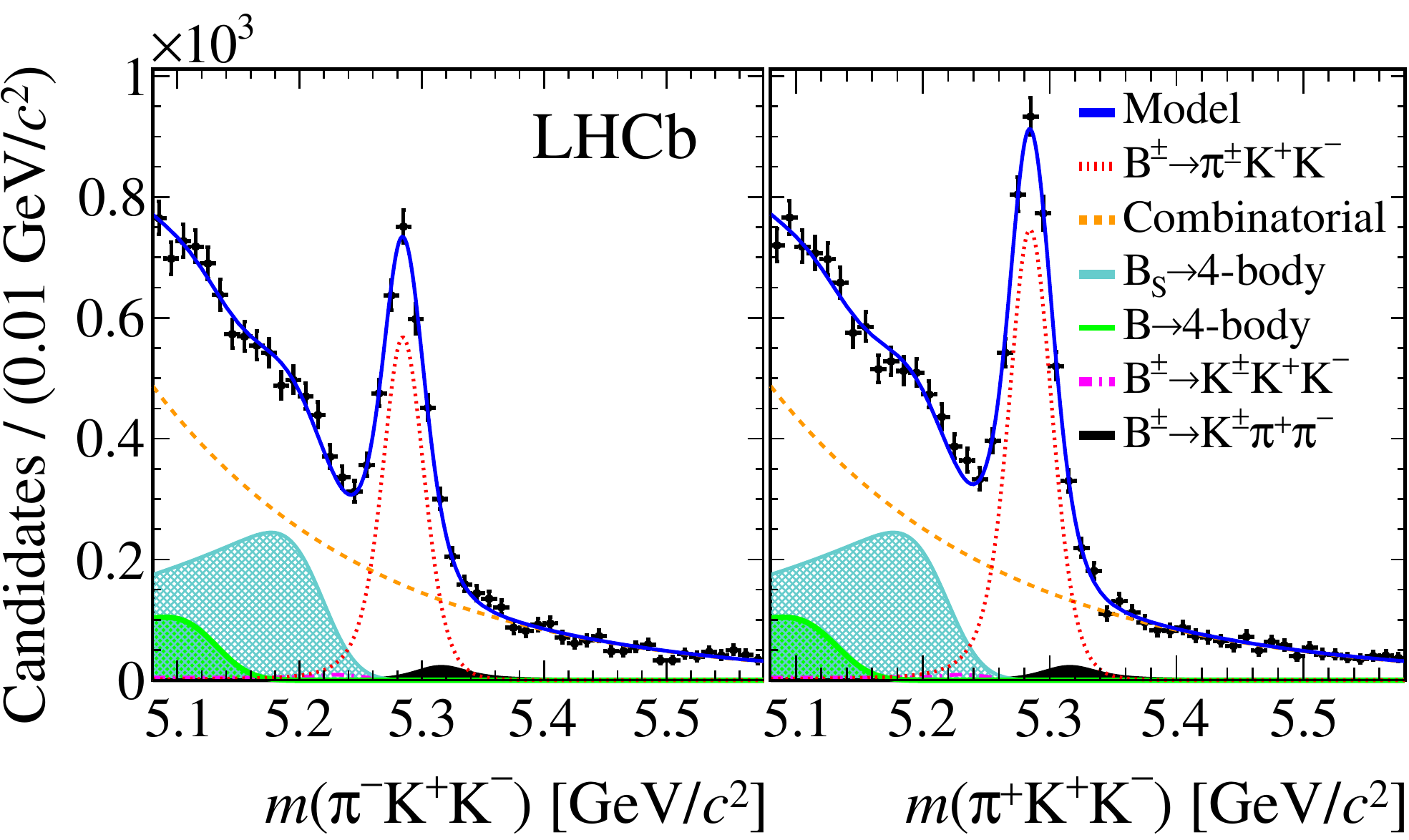}
  \put(15,45){\bf{(d)}}
\end{overpic}
\caption{Invariant mass spectra of (a) \kpipi, (b) \kkk, (c) \pipipi and (d) \kkpi decays.
The left panel on each figure shows the \Bm candidates and the right panel shows the \Bp candidates.
The results of the unbinned maximum likelihood fits are overlaid. The main components of the fits are also shown.
}
\label{fig:MassFit}
\end{figure}

The invariant mass spectra of the four decay modes are shown in Fig.~\ref{fig:MassFit}.
The figure is illustrative only, as the asymmetries are obtained from separate fits of the samples divided by year, trigger selection and magnet polarity, and then combined as described in Sec.~\ref{sec:acpmeasurement}. 

The signal yields obtained for the combined 2011 and 2012 data samples are shown in Table~\ref{tab:signalyields}.
The data samples are larger than those presented in Refs.~\cite{LHCb-PAPER-2013-027} and~\cite{LHCb-PAPER-2013-051} due to both an increase in the integrated luminosity and the use of a more efficient selection.

\begin{table}[htb]
\caption{Signal yields of charmless three-body \Bpm decays for the full data set.}
\centering
\begin{tabular}{lc}
\hline
Decay mode  & Yield \\ \midrule
\kpipi  & $181\,074 \pm 556 $ \\
\kkk    & $109\,240 \pm 354 $ \\
\pipipi & $\;\, 24\,907 \pm 222 $ \\
\kkpi   & $\;\;\,\,  6\,161 \pm 172 $ \\ \hline
\end{tabular}
\label{tab:signalyields}
\end{table}

\section{Inclusive \boldmath{\CP} asymmetry measurement}
\label{sec:acpmeasurement}

The \CP asymmetry of \Bpm decays to a final state $f^{\pm}$ is defined as
\begin{equation}
\acp \equiv \frac{ \Gamma[\Bm \to f^-]  -  \Gamma[\Bp\to f^+] }{ \Gamma[\Bm \to f^-]  + \Gamma[\Bp\to f^+] } \, ,
\label{eq:acpdef}
\end{equation}
where $\Gamma$ is the partial decay width.
To determine the inclusive \CP asymmetries, the raw asymmetries measured from the fits are corrected for effects induced by the detector efficiency, interactions of final-state particles with matter, and any asymmetry in the forward production rates between \Bp and \Bm mesons.
The raw asymmetry, \acpraw, is written in terms of the \Bm and \Bp event yields as
\begin{equation}
\acpraw \equiv \frac{N_{\Bm} -N_{\Bp}}{N_{\Bm} + N_{\Bp}} \, ,
\label{eq:arawdef}
\end{equation}
where the numbers of signal events $N_{B^-}$ and $N_{B^+}$ are related to the asymmetries by
\begin{eqnarray}
N_{B^-} &=& (1 + \acp + \adet + \aprod) \frac{N_S}{2} \times \frac{\langle \varepsilon^+\rangle}{\langle \varepsilon^-\rangle} \,\nonumber \\
N_{B^+} &=& (1 - \acp - \adet - \aprod) \frac{N_S}{2}.
\label{eq:yields}
\end{eqnarray}
Here, \aprod is the \Bpm  meson production asymmetry, $N_S$ are the total yields, and $\langle \varepsilon^{\pm}\rangle$ are the average efficiencies for selecting and reconstructing \Bp and \Bm decays, respectively.
The efficiency is computed on an event-by-event basis and depends on the position in the Dalitz plot.
The term \adet accounts for residual detection  asymmetries, such as differences in interactions of final state particles with the detector material or left-right asymmetries that may not be properly represented in the Monte Carlo.

The three final-state hadrons are treated as the combination of a pair of same-flavour, charge-conjugate hadrons $h^+h^-=\pip\pim,\Kp\Km$, and an unpaired hadron $h'^{\pm}$ with the same charge as the \Bpm meson.
The detection asymmetry $\adet^{h'}$ is given in terms of the charge-conjugate detection efficiencies  of the unpaired hadron $h'^{\pm}$, and
the production asymmetry \aprod is given in terms of the \Bpm production rates.

The raw asymmetry is expressed in terms of \acp, \aprod and $\adet^{h'}$ using Eqs.~(\ref{eq:arawdef}) and~(\ref{eq:yields}),
\begin{equation}
\acpraw = \frac{ \acp + \aprod + \adet^{h'} + \acp \aprod \adet^{h'}}{ 1 + \acp \aprod + \acp \adet^{h'} + \aprod \adet^{h'} } \, .
\label{eq:arawFull}
\end{equation}
For small asymmetries the products are negligible, and the raw asymmetry becomes
\begin{equation}
 \acpraw \approx \acp + \aprod + \adet^{h'} \, .
 \label{acpapprox}
\end{equation}
Throughout this paper, Eq.~(\ref{acpapprox}) is used in calculating the inclusive asymmetries, as all terms are sufficiently small.
For the determination of the asymmetries in regions of the phase space where the raw asymmetries are large, the full formula of Eq.~(\ref{eq:arawFull}) is applied.

The four decay channels are divided into two categories according to the flavour of the final-state hadron $h'^{\pm}$.
For the \kpipi and \kkk decay channels, the \CP asymmetry is expressed in terms of the raw asymmetry and  correction terms given by the sum of the \Bpm production asymmetry and the kaon detection asymmetry, \aprod and \adetk.
For the \kkpi and \pipipi decay channels, the pion detection asymmetry \adetpi is used. The \CP asymmetries are calculated as
\begin{eqnarray}
\label{eq:acpkpipikkk-acppipipikkpi}
\acp(Khh) &= & \acpraw(hhK) - \aprod - \adetk = \acpraw(hhK) - A_{\Delta}\, ,\nonumber \\
\acp(\pi hh) &= & \acpraw(hh\pi) - \aprod - \adet^{\pi} = \acpraw(hh\pi) - A_{\Delta} + \adetk -
 \adet^{\pi}\,  . 
\end{eqnarray}
The correction term $A_{\Delta}$ is measured using approximately $265\,000$ $\Bpm \to \jpsi (\mup\mu^-)\Kpm$ decays. The correction is obtained from the raw asymmetry of the \jpsik mode as
\begin{equation}
A_{\Delta} = \acpraw(\jpsi \Kpm) - \acp(\jpsi \Kpm)  , 
\label{deltaJpsik}
\end{equation}
using the world average of the \CP asymmetry  $\acp(\jpsi \Kpm) = (0.1\pm 0.7)\%$~\cite{PDG2012}.

The pion detection asymmetry, $\adetpi =(0.00\pm0.25)\%$, has been previously measured by LHCb~\cite{LHCb-PAPER-2012-009} and is consistent with being independent of \ptot and \pt.
The production asymmetry is obtained from the same sample of \jpsik decays as $\aprod = A_{\Delta} - \adetk$, and is consistent with being constant in the interval of momentum measured.
Here the kaon interaction asymmetry $\adetk=(-1.26\pm0.18)\%$ is measured in a sample of  $\Dstarp\to\pip \Dz\to \pip \Km\pip\pim\pip$ decays, where the $\Dstarp$ is produced in the
decay of a $B$ meson. The value of $\adetk$ is obtained
by measuring the ratio of fully to partially reconstructed $\Dstarp$ decays~\cite{LHCb-PAPER-2012-009}.

Since neither the detector efficiencies nor the observed raw asymmetries are uniform across the Dalitz plot,
an acceptance correction is applied to the integrated raw asymmetries.
This is determined by the ratio of the \Bm and \Bp average efficiencies in simulated events, reweighted to reproduce the population of signal data in bins of the Dalitz plot.
In addition, to account for the small charge asymmetry introduced by the hadronic hardware trigger, the data are divided into the trigger independent of signal and the trigger on signal samples, as discussed in Sec.~2.
The \CP asymmetries are calculated using Eqs.~(\ref{eq:acpkpipikkk-acppipipikkpi}) and (\ref{deltaJpsik}), applied to the acceptance-corrected raw asymmetries of the samples collected in each trigger configuration.
The inclusive \CP asymmetry of each mode is the weighted average of the \CP asymmetries for the samples divided by trigger and year of data taking, taking into account the correlation between trigger samples as described in Ref.~\cite{Lyons1988110}.

\section{Systematic uncertainties and results}
\label{sec:syst}
Several sources of systematic uncertainty are considered. These include potential mismodellings in the mass fits, the phase-space acceptance corrections and the trigger composition of the samples.

The systematic uncertainties due to the mass fit models are evaluated as the full difference in \CP asymmetry resulting from variations of the model.
The alternative fits have good quality and describe the data accurately.
To estimate the uncertainty due to the choice of the signal mass function, the initial model is replaced by an alternative empirical distribution~\cite{Santos:2013gra}.
A systematic uncertainty to account for the use of equal means and widths for \Bm and \Bp signal peaks in the default fit is assigned by repeating the fits with these parameters allowed to vary independently.
The resulting means and widths are found to agree and the difference in the value of \acp is assigned as a systematic uncertainty. 

The systematic uncertainty associated with the peaking background fractions reflects the uncertainties in the expected yields determined from simulation, and the influence of combining 2011 and 2012 simulated samples when determining the fractions in the nominal fit, by repeating the fits with the background fractions obtained for the samples separately.
The uncertainty due to background shape is obtained by increasing the width of the Gaussian function according to the observed differences between simulation and data for peaking backgrounds, and allowing the four-body shape to vary in the fit.
Similarly, the possibility of non-zero background asymmetries is tested by letting the peaking and four-body-background normalisations vary separately for \Bm and \Bp fits.
The signal model variations and the background asymmetry are the dominant systematic uncertainties related to the fit procedure.

The systematic uncertainty related to the acceptance correction procedure consists of two parts:  the statistical uncertainty on the detection efficiency due to the finite size of the simulated samples,
and the uncertainty due to the choice of binning, which is evaluated by varying the binning used in the efficiency correction.

A study is performed to investigate the effect of having different trigger admixtures in  the signal and the control channels.
The acceptance-corrected \CP asymmetries are measured separately for each trigger category  and found to agree, therefore no additional systematic uncertainty is assigned.
Performing this comparison validates the assumption that the detection asymmetry factorises between the $h^+h^-$ pair and the $h'^{\pm}$, within the statistical precision of the test.

\begin{table}[tb]
\centering
\caption{Systematic uncertainties on the measured asymmetries, where the total is the sum in quadrature of the individual contributions. The \adetpi uncertainty is taken from Ref.~\cite{LHCb-PAPER-2012-009}.
}
\begin{tabular} {l cc cc cc cc}
\hline
Systematic  & \multicolumn{2}{c}{$\acp(K\pi\pi)$}  & \multicolumn{2}{c}{$\acp(KKK)$} &  \multicolumn{2}{c}{$\acp(\pi\pi\pi)$}  &  \multicolumn{2}{c}{$\acp(\pi KK)$}     \\
uncertainty     &   2011 &2012   &   2011 &2012  &   2011 &2012  &   2011 &2012                         \\
\hline
Signal model    &  &  & &  &  &  &  &  \\ 
\; function & 0.0000 & 0.0001 & 0.0002 & 0.0005 & 0.0046 & 0.0025 & 0.0028 & 0.0046 \\ 
\; \Bpm parameters      & 0.0006 & 0.0006 & 0.0005 & 0.0005 & 0.0032 & 0.0032 & 0.0027 & 0.0027 \\
Background &  &  & &  &  &  &  &  \\
\; fractions    & 0.0000 & 0.0001 & 0.0002 & 0.0001 & 0.0001 & 0.0001 & 0.0010 & 0.0014 \\ 
\; resolution           & 0.0000 & 0.0001 & 0.0001 & 0.0000 & 0.0001 & 0.0000 & 0.0001 & 0.0004 \\ 
\; asymmetry    & 0.0031 & 0.0032 & 0.0015 & 0.0011 & 0.0017 & 0.0027 & 0.0011 & 0.0019 \\ 
Acceptance corr. &  0.0012 &  0.0018 &0.0013 &0.0013 &0.0063 & 0.0051 &  0.0099 & 0.0092 \\
\adetk uncertainty & -- & -- & -- & -- & 0.0018 & 0.0018 & 0.0018 & 0.0018 \\
\adetpi uncertainty & -- & -- & -- & -- & 0.0025 & 0.0025 & 0.0025 & 0.0025 \\
\hline
Total &0.0034 & 0.0038 & 0.0020 & 0.0019 & 0.0090&  0.0075 & 0.0113 &0.0115 \\
\hline
\end{tabular}
\label{tab:sysperyear}
\end{table}

The systematic uncertainties, separated by year, are shown in Table~\ref{tab:sysperyear}, where the total systematic uncertainty is  the sum in quadrature of the individual contributions.
The uncertainties on \adetpi and \adetk are only considered as systematic uncertainties for \pipipi and \kkpi decays, following Eq.~(\ref{eq:acpkpipikkk-acppipipikkpi}).
The systematic uncertainty of the 2011 and 2012 combination is taken to be the greater of these two values.

The results for the integrated \CP asymmetries are
\begin{eqnarray}
\acp(\kpipi)&=& +\kpipiacp \, , \nonumber \\
\acp(\kkk) &=& \kkkacp \, , \nonumber \\
\acp(\pipipi)&=& +\pipipiacp \, , \nonumber \\
\acp(\kkpi)&=& \kkpiacp \, , \nonumber
\end{eqnarray}
where the first uncertainty is statistical, the second systematic, and the third is due to the limited knowledge of the \CP asymmetry of the \jpsik reference mode~\cite{PDG2012}.
The significances of the inclusive charge asymmetries, calculated by dividing the central values by the sum in quadrature of the uncertainties, are $\kpipisigma$ standard deviations ($\sigma$) for \kpipi decays, $\kkksigma\sigma$ for \kkk decays, $\pipipisigma\sigma$ for \pipipi decays and $\kkpisigma\sigma$ for \kkpi decays.

\section{\boldmath{\CP} asymmetry in the phase space}

\begin{figure}[tbp!]
\begin{center}
\begin{overpic}[width=0.49\linewidth]{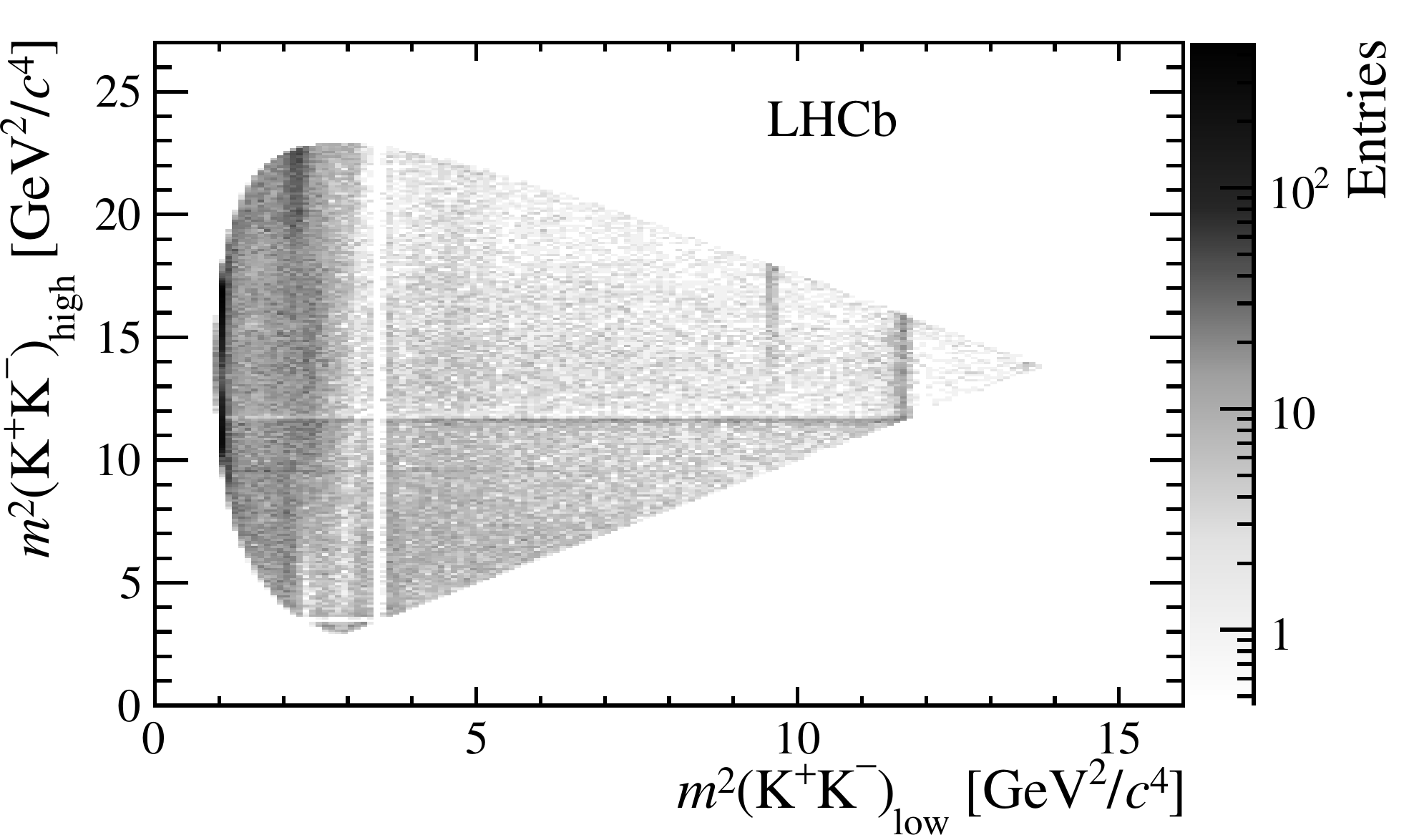}
  \put(73,50){\bf{(a)}}
\end{overpic}
\begin{overpic}[width=0.49\linewidth]{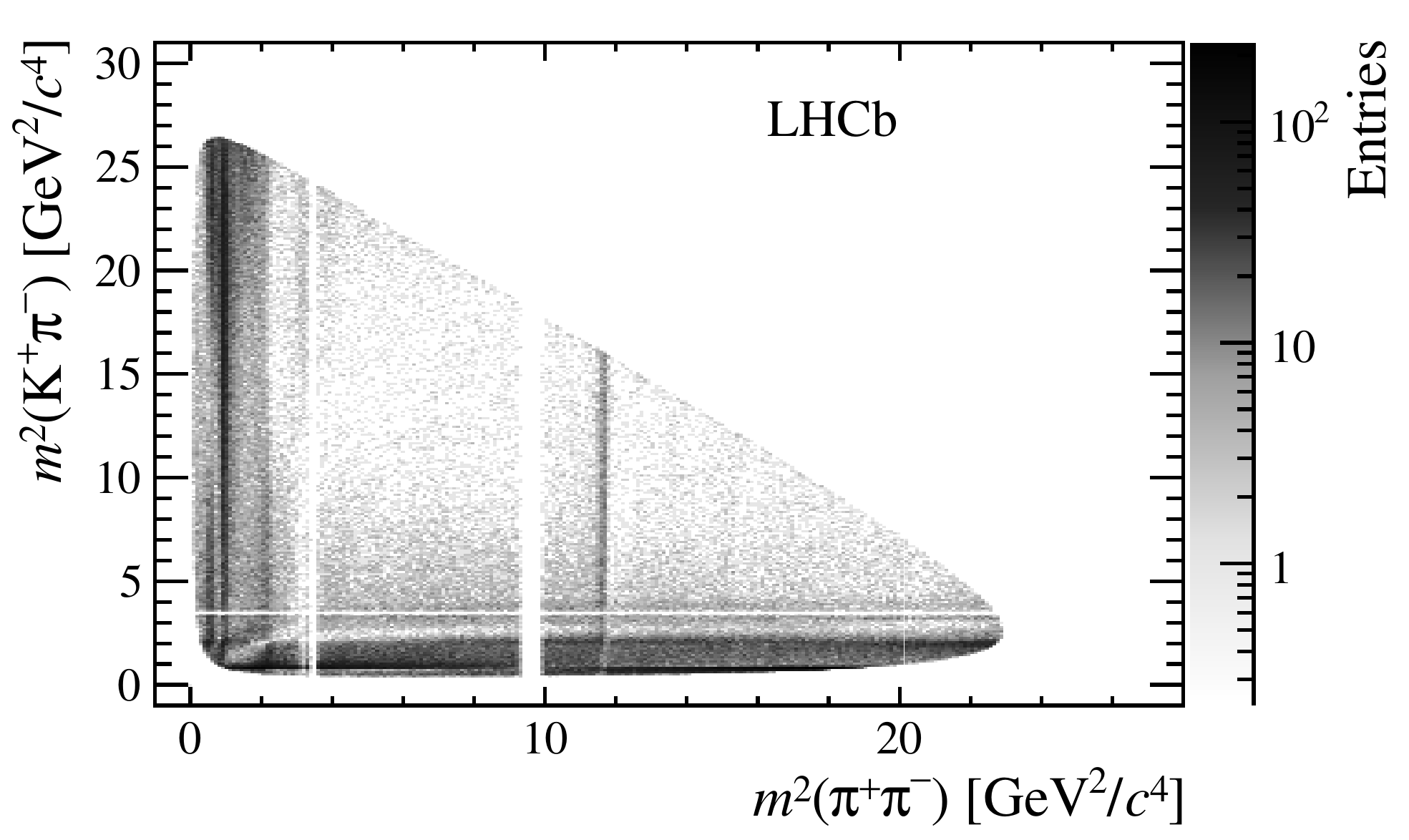}
  \put(73,50){\bf{(b)}}
\end{overpic}
\begin{overpic}[width=0.49\linewidth]{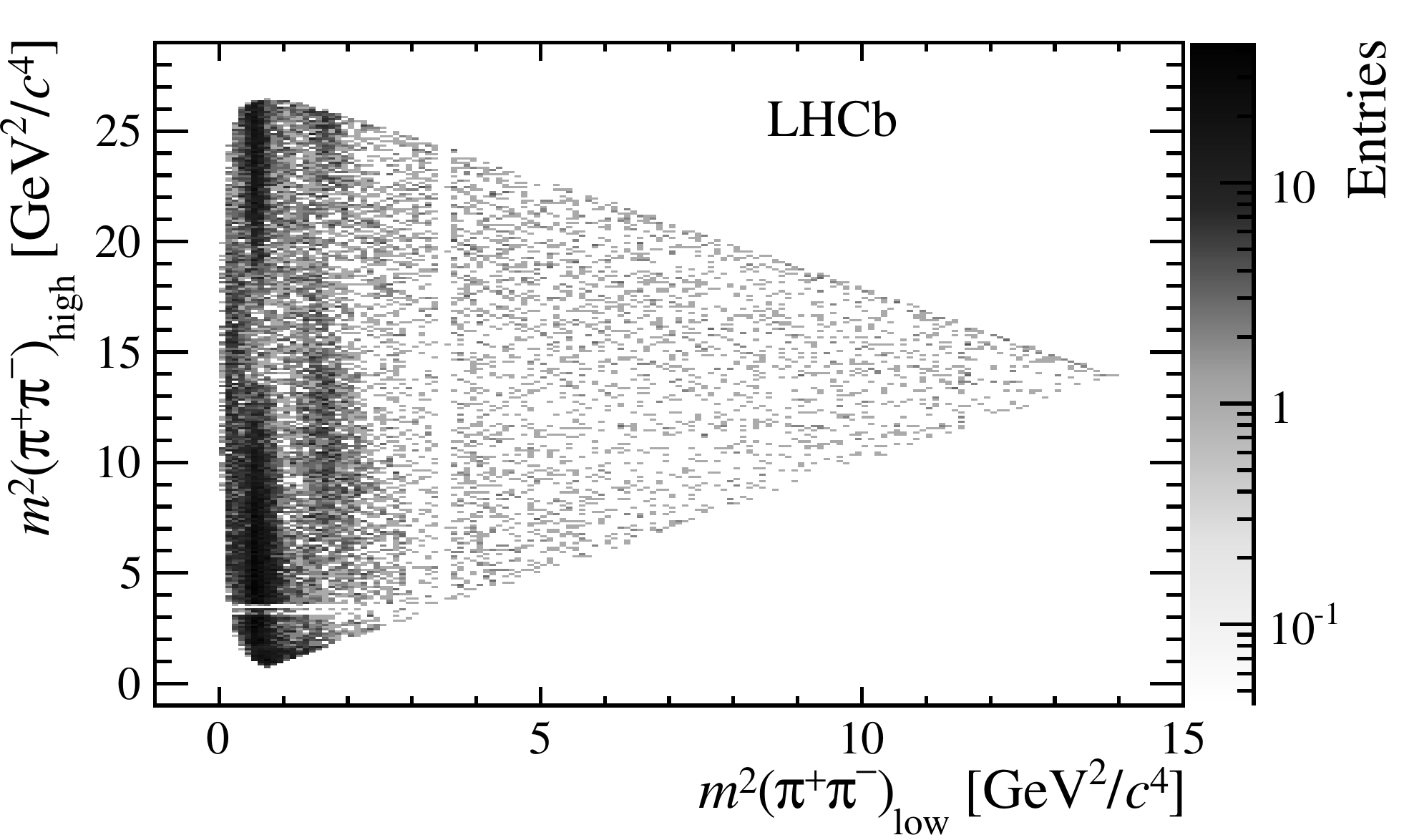}
  \put(73,50){\bf{(c)}}
\end{overpic}
\begin{overpic}[width=0.49\linewidth]{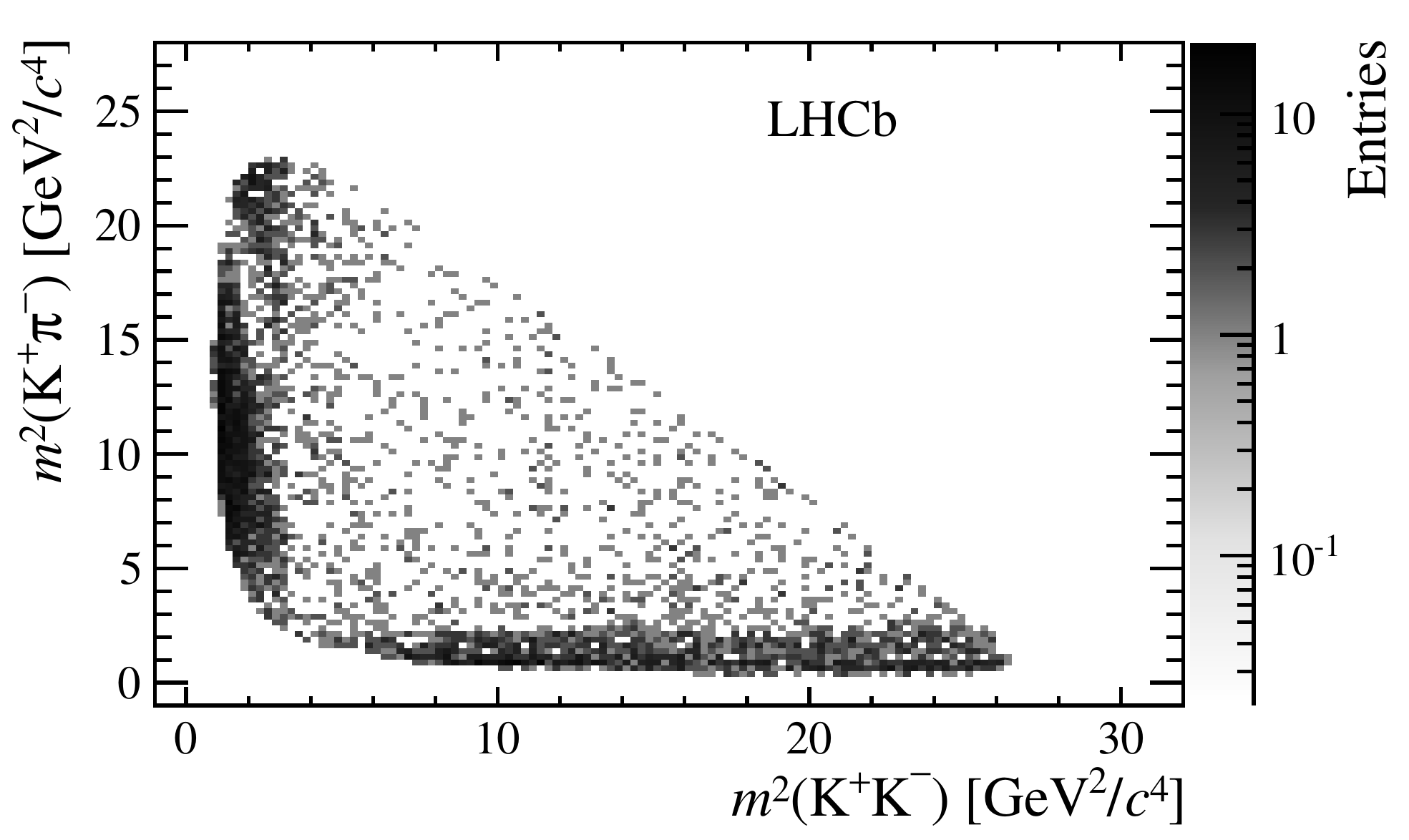}
  \put(73,50){\bf{(d)}}
\end{overpic}
\caption{Dalitz plot distributions of (a)  \kkk, (b) \kpipi, (c) \pipipi and (d) \kkpi candidates. The visible gaps correspond to the exclusion of the $J/\psi$ (in the \kpipi decay) and \Dz (all plots, except for the \kkpi decay) mesons from the samples.}
\label{fig:DP_all}
\end{center}
\end{figure}

The Dalitz plots distributions in the signal region  for the four channels are shown in Fig.~\ref{fig:DP_all}.
For the \kkk and \pipipi decays, folded Dalitz plots are used.
For a given event, the vertical axis of the Dalitz plot corresponds to the invariant mass squared of the decay with the highest value ($m^2(h^+h^-)_{\text{high}}$), while the horizontal axis is the invariant mass squared with the lowest value between the two ($m^2(h^+h^-)_{\text{low}}$).

The signal region is defined as the three-body invariant mass region within $34 \mevcc$ of the fitted mass, except for the \kkpi channel,  for which the mass window is restricted to $\pm 17\mevcc$ of the peak due to the larger background. The expected background contribution is not subtracted from the data presented in these figures. To improve the resolution, the Dalitz variables are calculated after refitting the candidates with their invariant masses constrained to the known \Bpm value~\cite{PDG2012}.
The events are concentrated in low-mass regions, as expected for charmless decays dominated by resonant contributions.

In the \kkk decays, the region of $\mmkklow$ around $1.0\gevgevcccc$ corresponds to the $\phi(1020)$ resonance, and that around $11.5\gevgevcccc$ to the $\chi_{c0}(1P)$ meson. In the region $2$--$3\gevgevcccc$, there are clusters that could correspond to the $f'_2(1525)$ or the $f_0(1500)$ resonances observed by BaBar in this decay mode~\cite{BaBarkkk}.
The contribution of \jpsik decays with $\jpsi \to K^+ K^-$ is visible around $9.6\gevgevcccc$ in $\mmkk$.

In the \kpipi Dalitz plot, there are low-mass resonances in both $\Kpm\pimp$ and  $\pi^+ \pi^-$ spectra: $\Kstarz(892)$, $\rho^0(770)$, $f_0(980)$ and $K^{*0}_{0,2}(1430)$.
In addition, the $\chi_{c0}(1P)$ resonance is seen at $\mmpipi \approx 11\gevgevcccc$.

For \pipipi decays, the resonances are the $\rho^0(770)$  at $\mmpipilow<1\gevgevcccc$.
In the region of $1.5 < \mmpipilow <2\gevgevcccc$, there are clusters that could correspond to the $\rho^0(1450)$, the $f_2(1270)$ and the $f_0(1370)$ resonances observed by BaBar in this decay mode~\cite{BaBarpipipi}.

For \kkpi decays, there is a cluster of events at $\mmkpi<2\gevgevcccc$, which could correspond to the $\Kstarz(892)$ and $K^{*0}_{0,2}(1430)$ resonances.
The \kkpi decays are not expected to have a contribution from the $\phi(1020)$ resonance~\cite{phiBR} and indeed, the $\phi(1020)$ contribution is not immediately apparent in the region of $\mmkk$ around $1\gevgevcccc$.

An inspection of the distribution of candidates from the $\Bp$ mass sidebands confirms that the background is not uniformly distributed, with combinatorial background events tending to be concentrated at the corners of the phase space, as these are dominated by low-momentum particles.

\begin{figure} [tb]
\begin{center}
\begin{overpic}[width=0.49\linewidth]{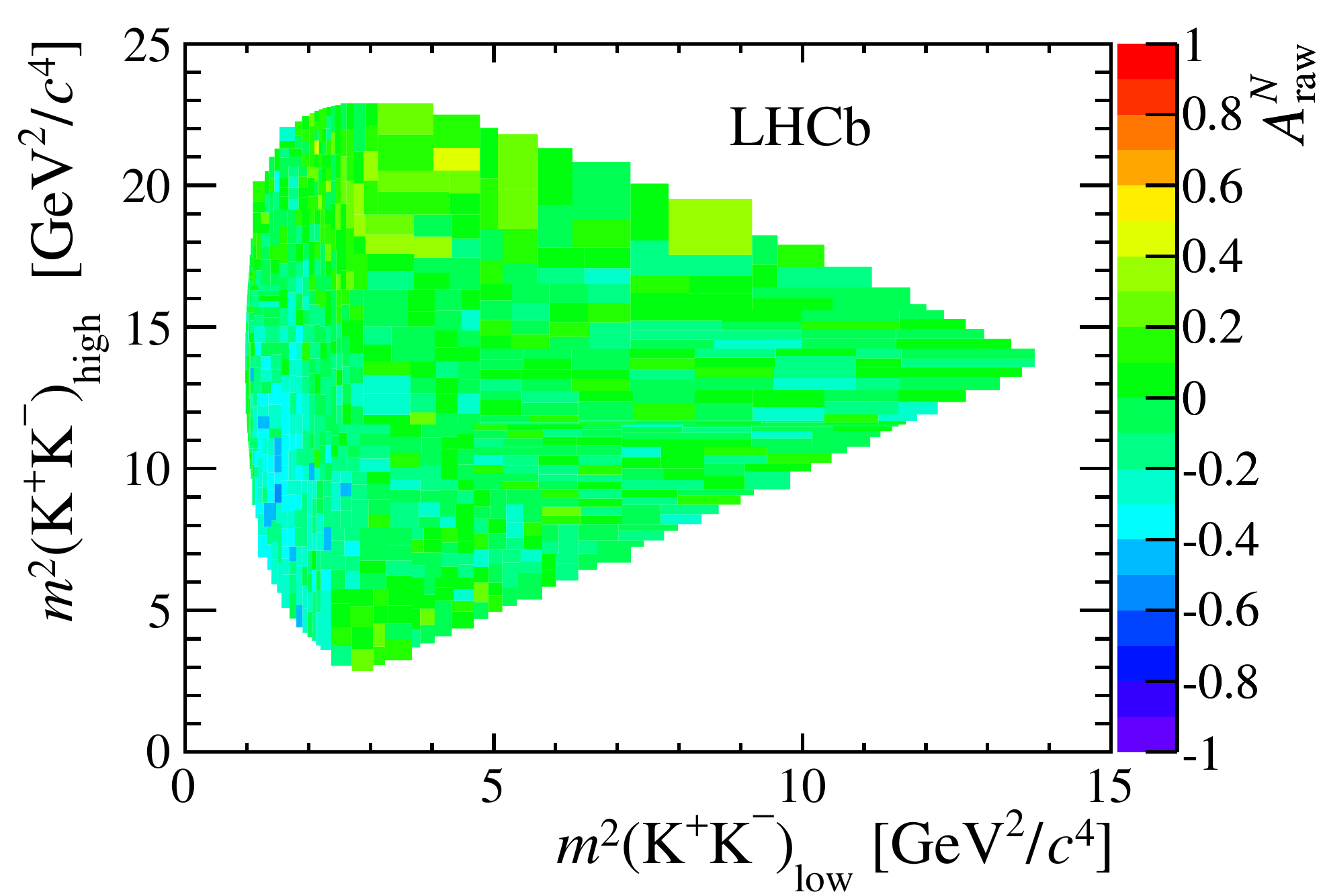}
 \put(73,58){\bf{(a)}}
\end{overpic}
\begin{overpic}[width=0.49\linewidth]{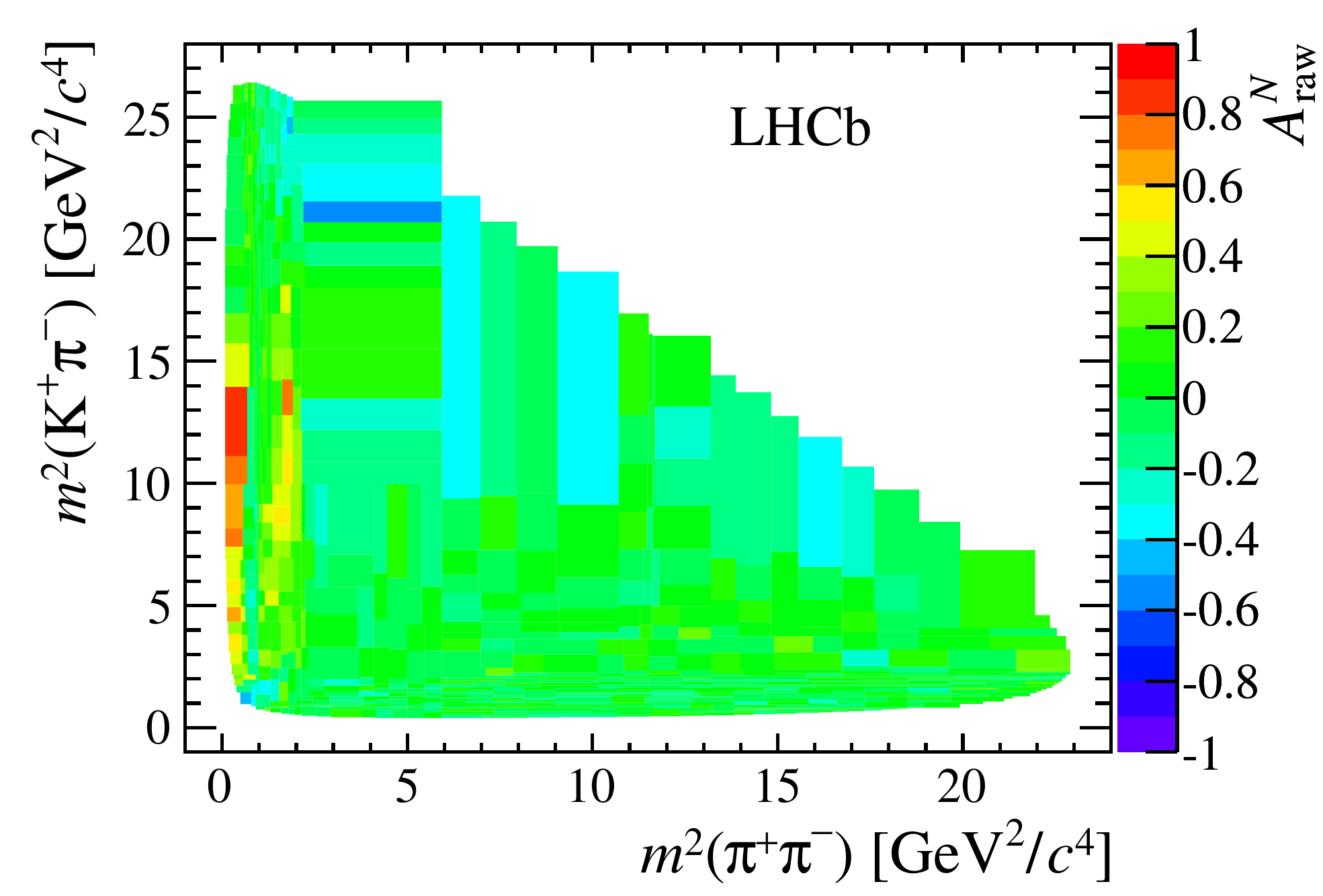}
 \put(73,58){\bf{(b)}}
\end{overpic}
\begin{overpic}[width=0.49\linewidth]{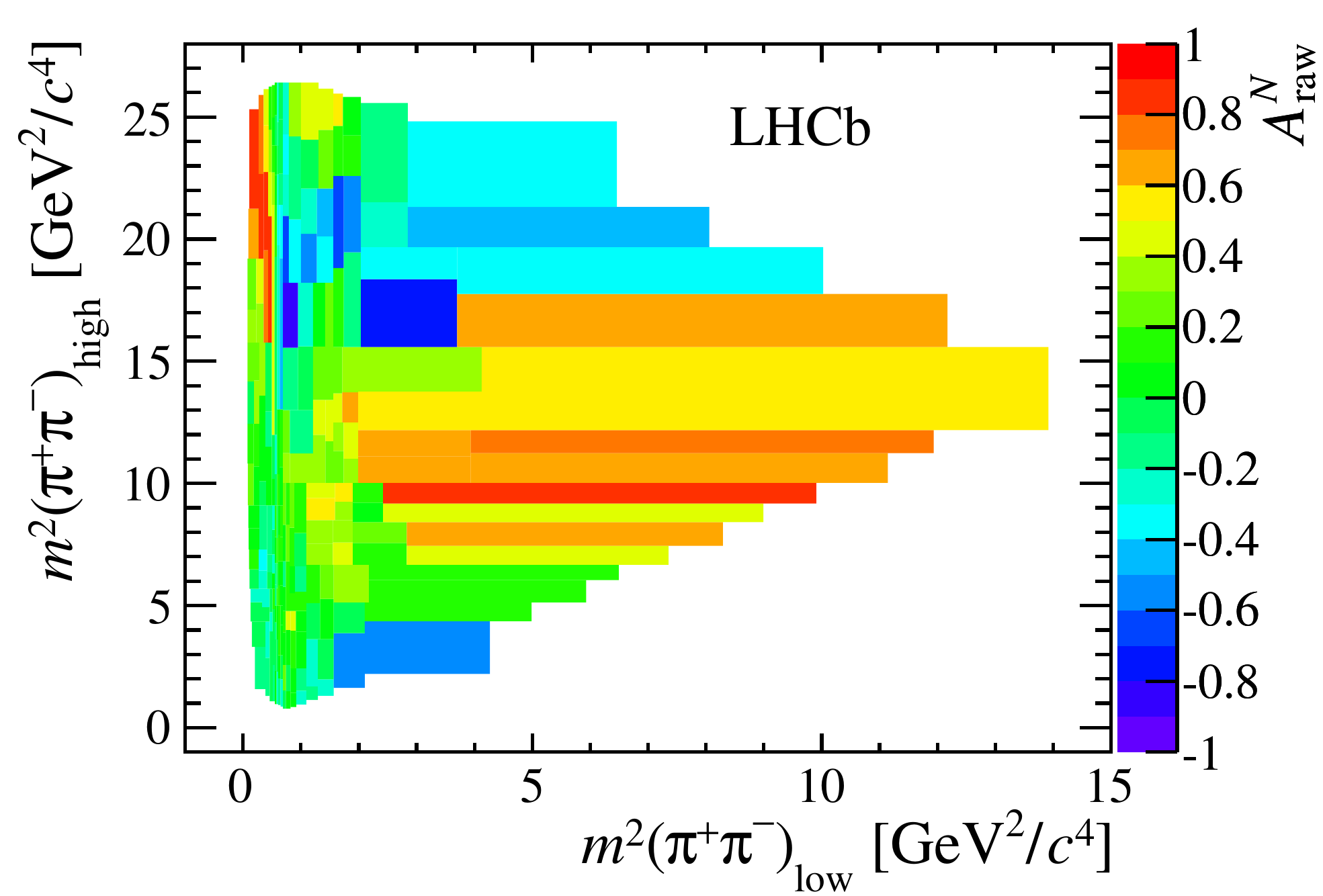}
 \put(73,58){\bf{(c)}}
\end{overpic}
\begin{overpic}[width=0.49\linewidth]{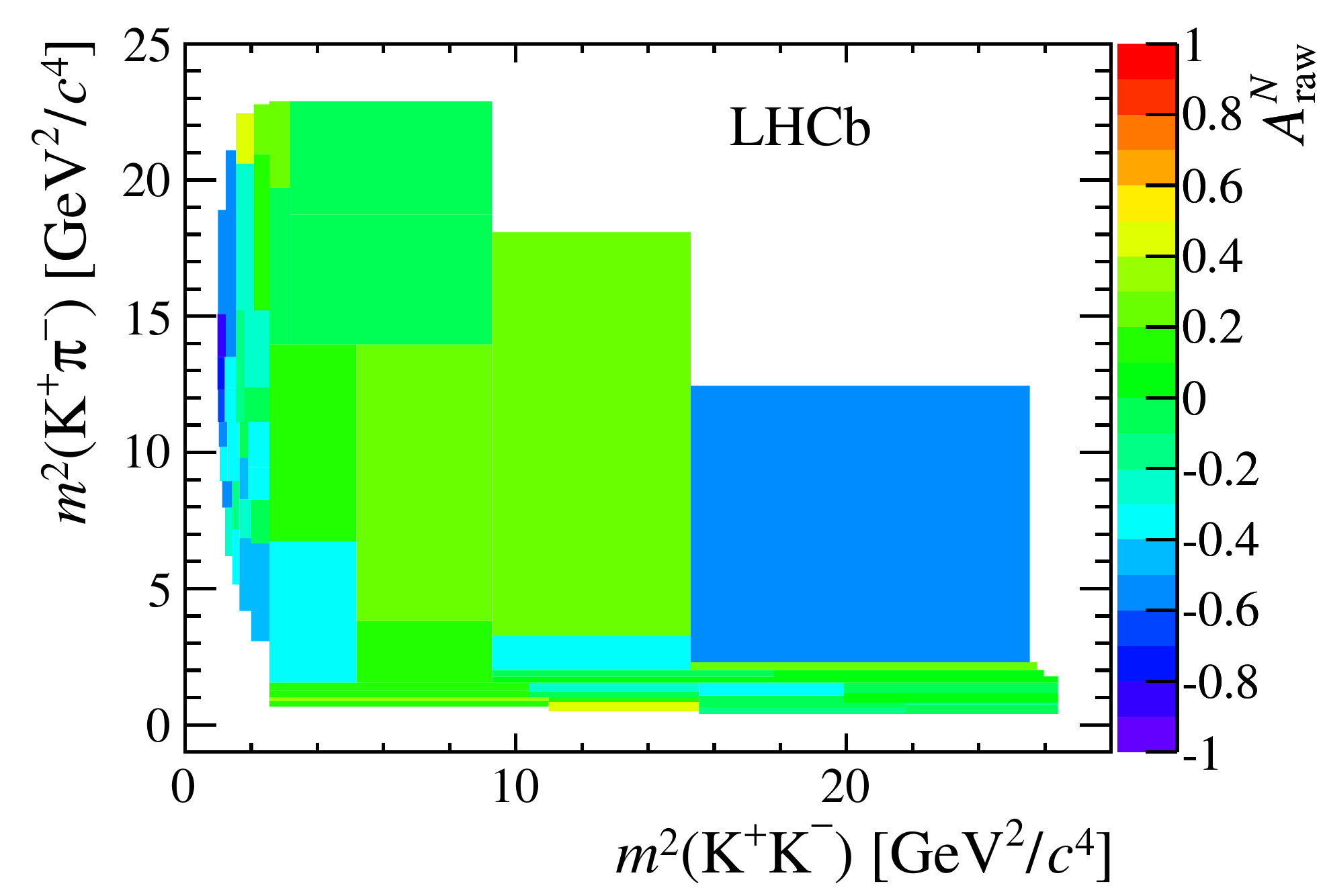}
 \put(73,58){\bf{(d)}}
\end{overpic}
\caption{(colour online)
Measured \acpn in Dalitz plot bins of background-subtracted and acceptance-corrected events for  (a) \kkk,  (b) \kpipi, (c) \pipipi and (d) \kkpi decays.
}
\label{fig:AdaptiveACP}
\end{center}
\end{figure}

In addition to the inclusive charge asymmetries, the asymmetries are studied in  bins of the Dalitz plots.
Figure~\ref{fig:AdaptiveACP} shows these  asymmetries, $\acpn \equiv  \frac{N^- - N^+}{N^- + N^+}$, where $N^-$ and $N^+$ are the background-subtracted, efficiency-corrected signal yields for $B^-$ and $B^+$ decays, respectively.
Background subtraction is done via a statistical tool to unfold data distributions called \sPlot\@ technique~\cite{Pivk:2004ty} using the \Bpm candidate invariant mass as discriminating variable.
The binning is chosen adaptively, to allow approximately equal populations of the total number of entries $(N^{-} + N^{+})$ in each bin.

\begin{figure}[tb]
\begin{center}
\begin{overpic}[width=0.48\linewidth]{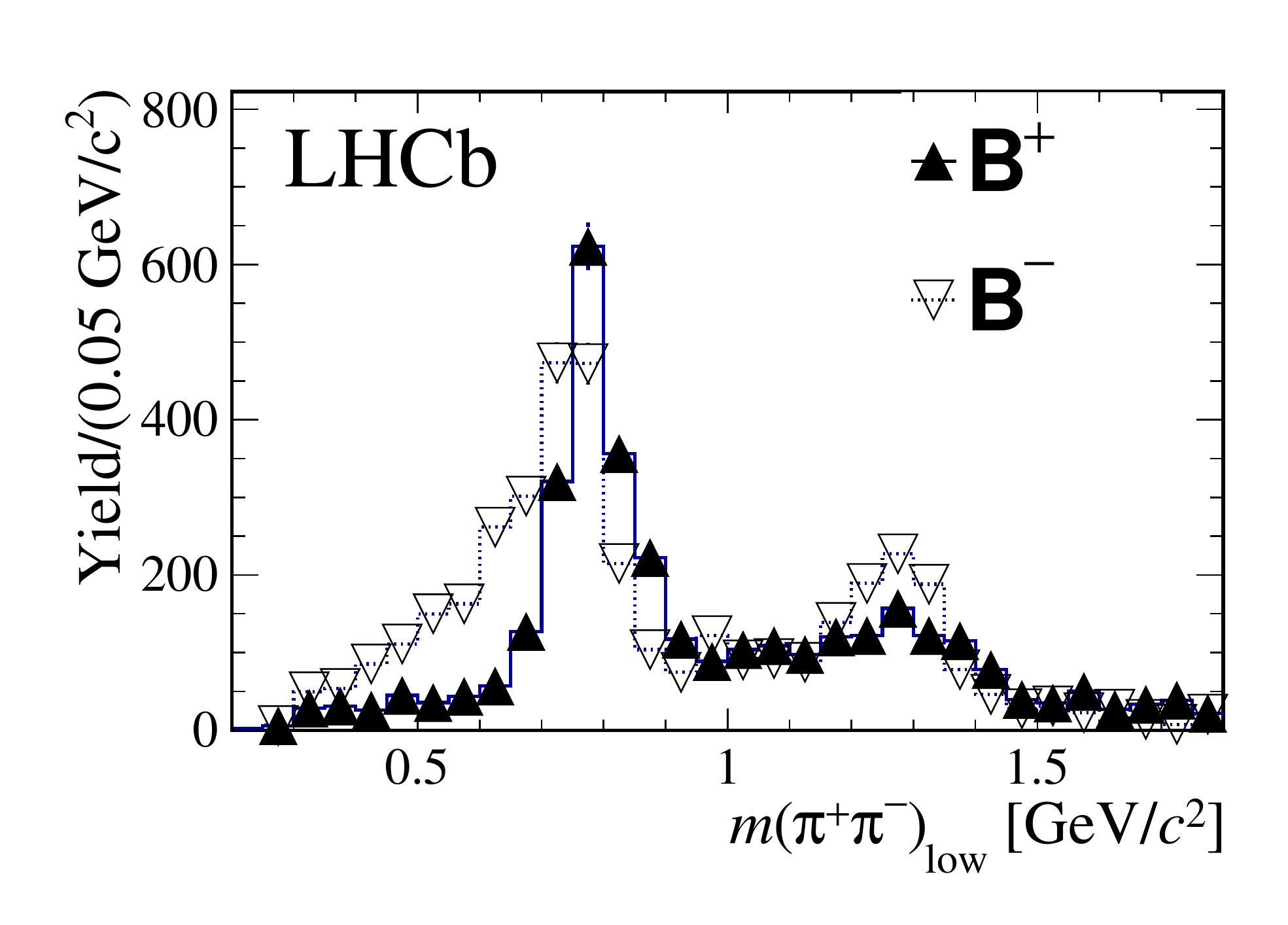}
 \put(85,58){\bf{(a)}}
\end{overpic}
\begin{overpic}[width=0.48\linewidth]{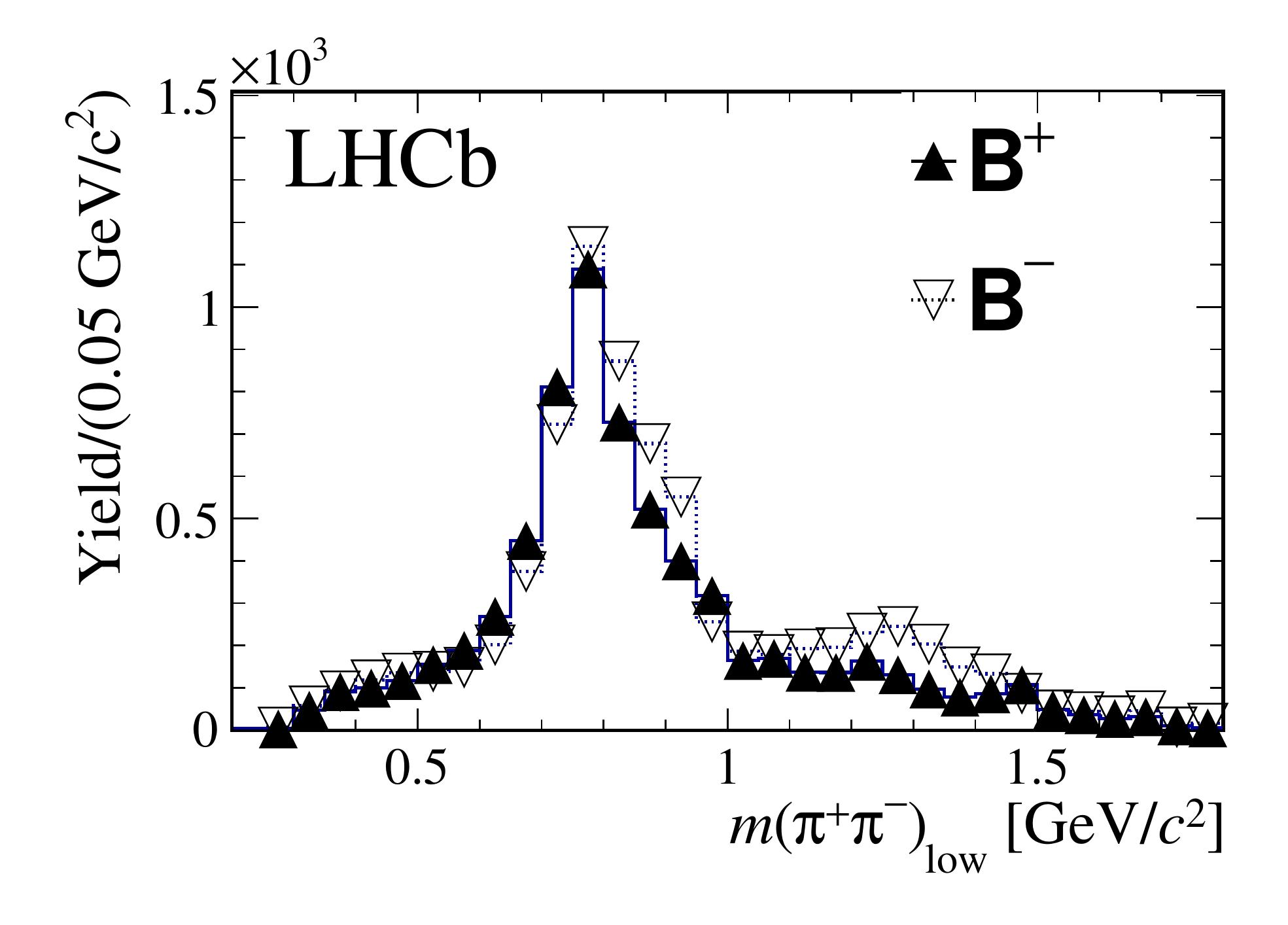}
  \put(85,58){\bf{(b)}}
\end{overpic}
\begin{overpic}[width=0.48\linewidth]{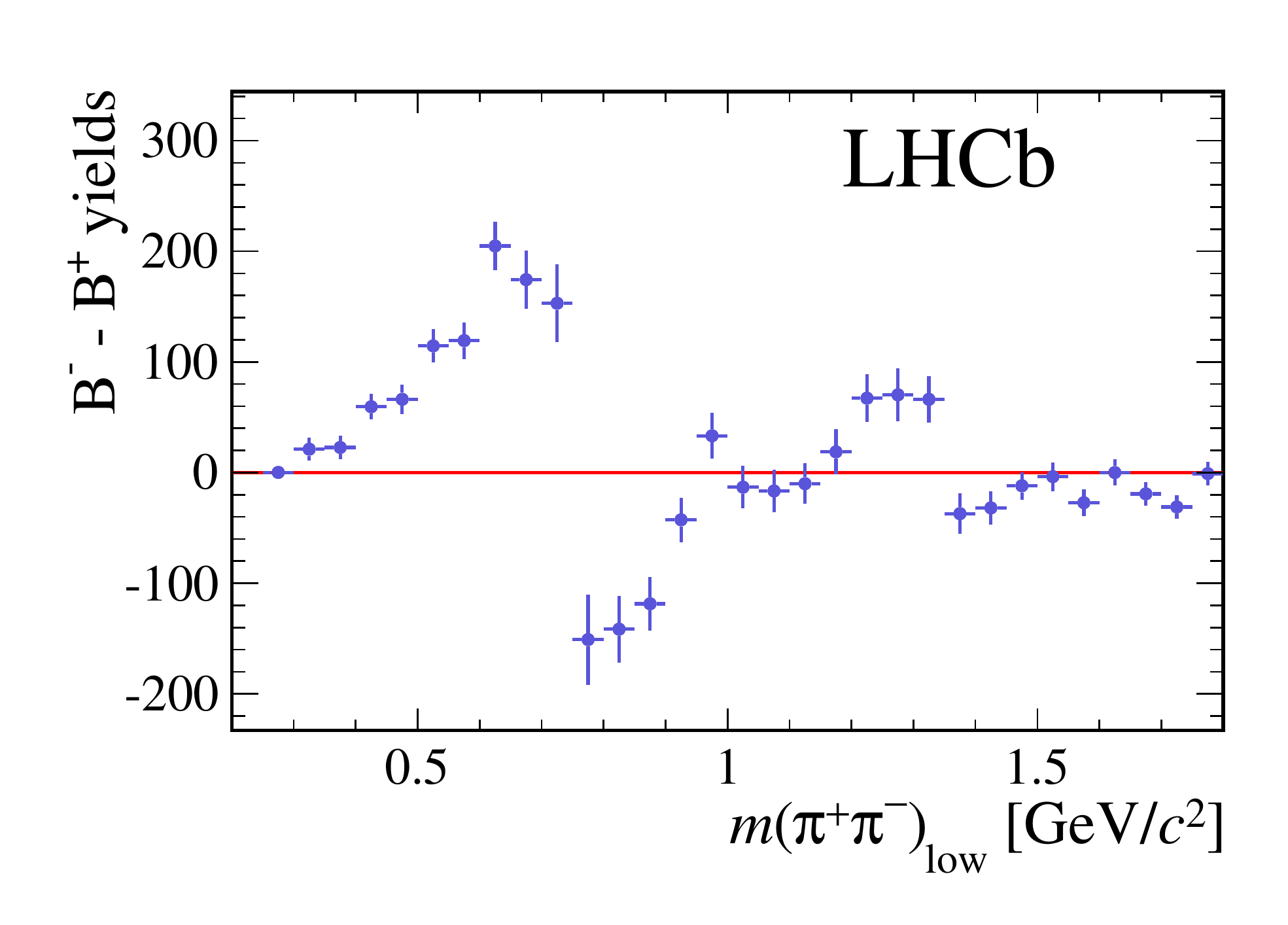}
  \put(85,58){\bf{(c)}}
\end{overpic}
\begin{overpic}[width=0.48\linewidth]{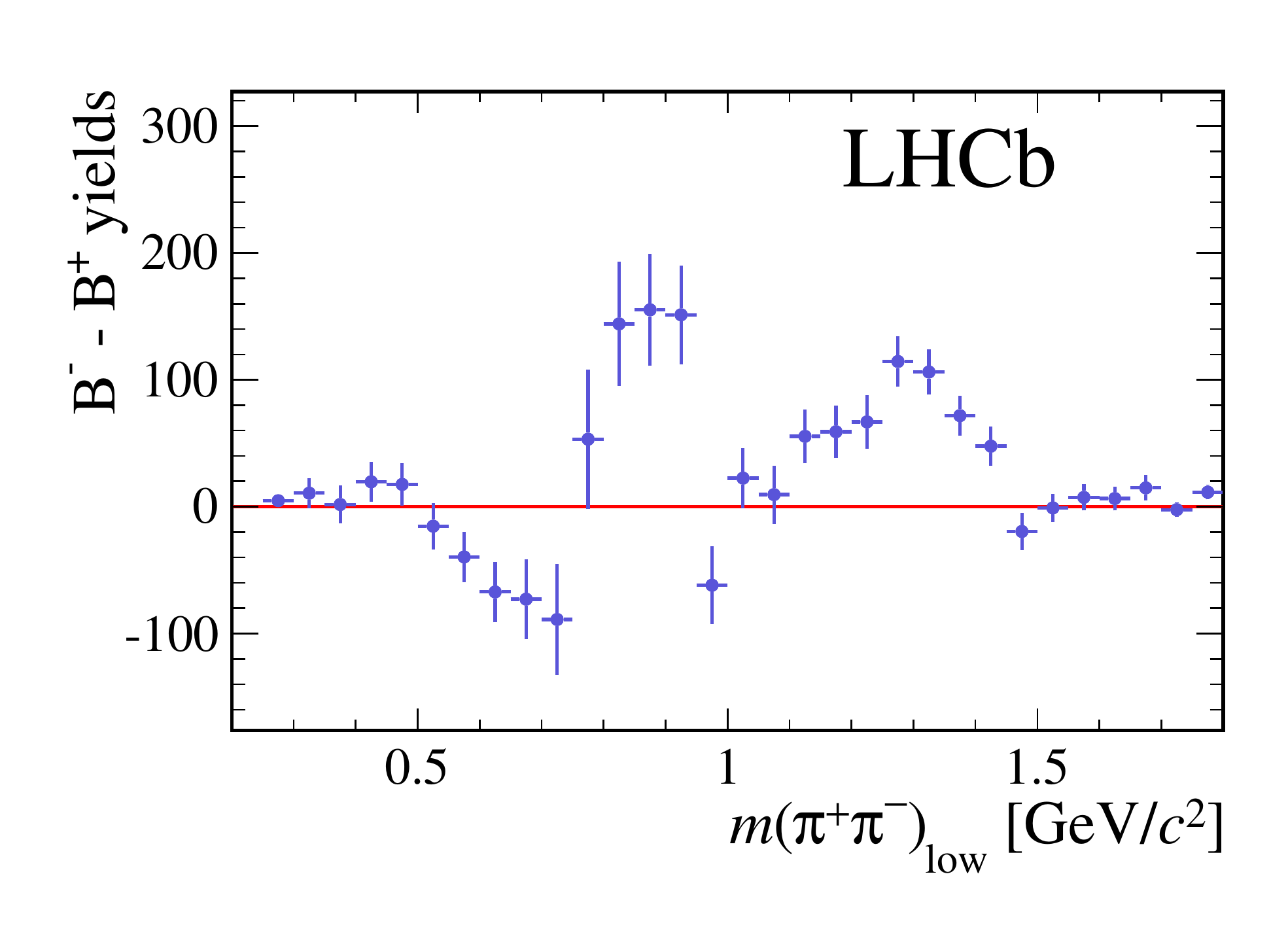}
  \put(85,58){\bf{(d)}}
\end{overpic}
\caption{
Projections in bins of the \mpipilow variable of (a, b) the number of \Bm and \Bp signal events and (c, d) their difference for \pipipi decays.
The plots are restricted to events with (a, c) $\cos\theta<0$  and (b, d) $\cos\theta>0$, with $\cos\theta$ defined in the text. The yields are acceptance-corrected and background-subtracted.
A guide line for zero (horizontal red line) was included on plots (c, d).
}
\label{fig:massProjpipipi:rho}
\end{center}
\end{figure}

\begin{figure}[tb]
\begin{center}
\begin{overpic}[width=0.48\linewidth]{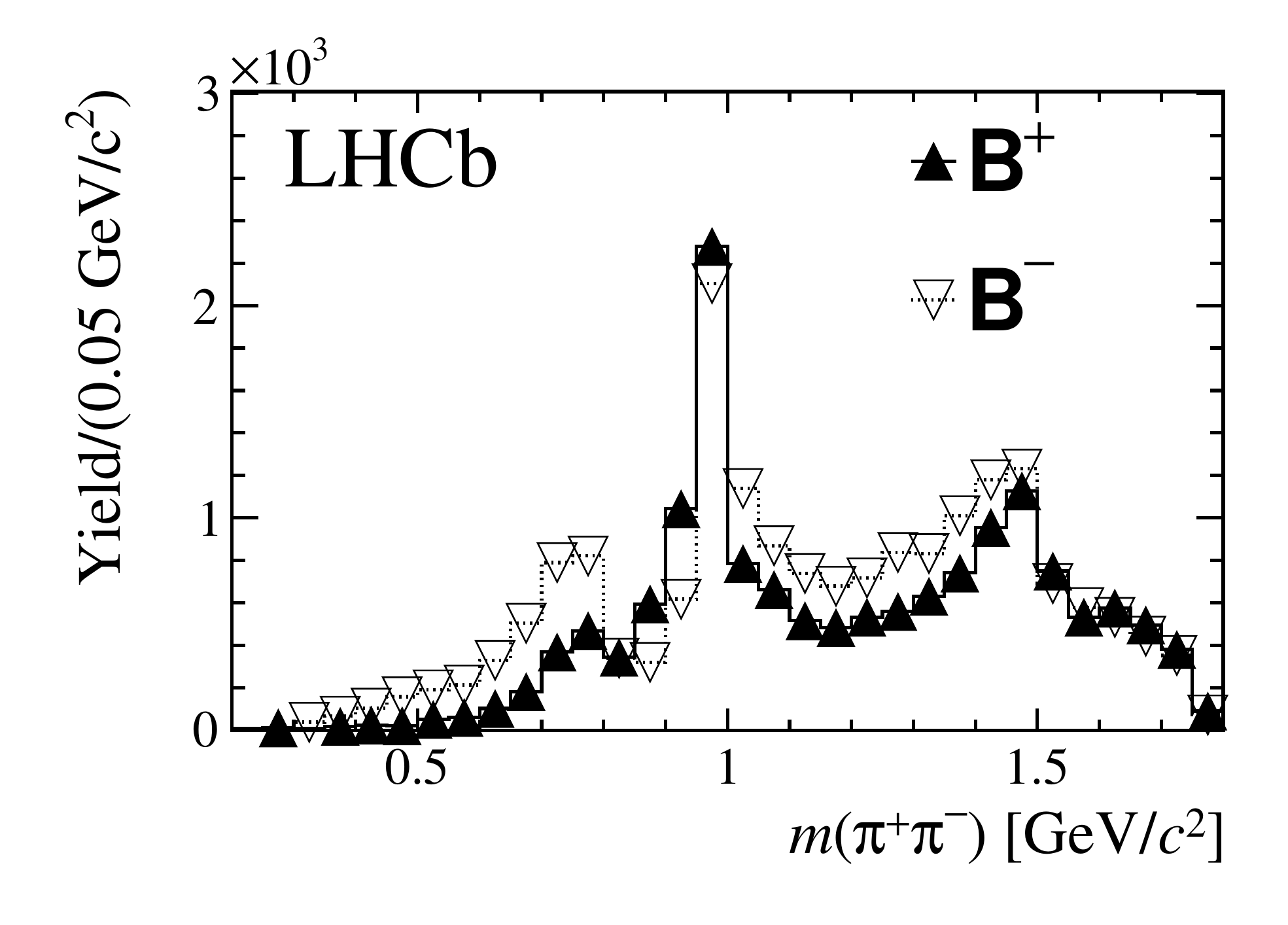}
 \put(85,58){\bf{(a)}}
\end{overpic}
\begin{overpic}[width=0.48\linewidth]{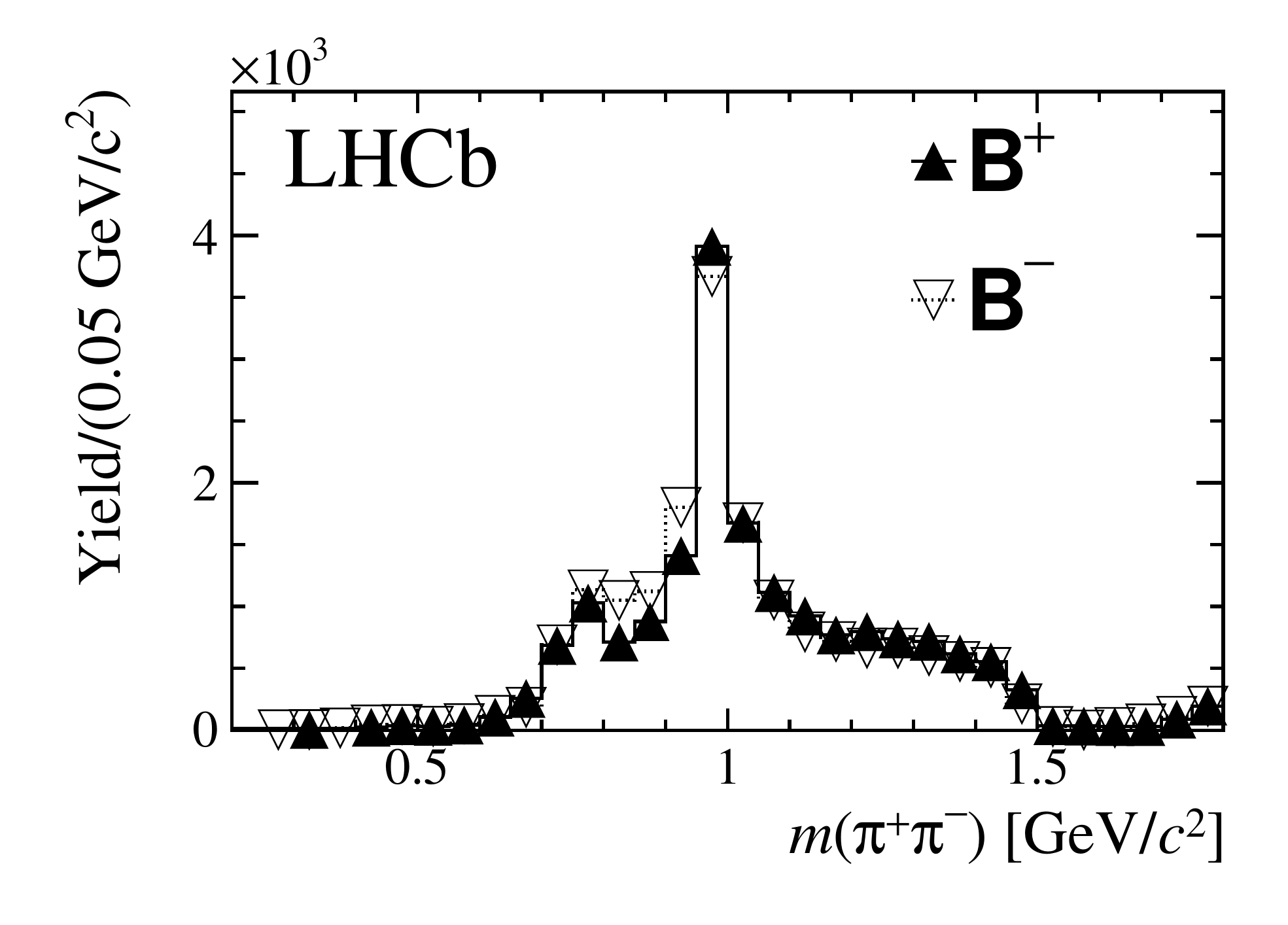}
 \put(85,58){\bf{(b)}}
\end{overpic}
\begin{overpic}[width=0.48\linewidth]{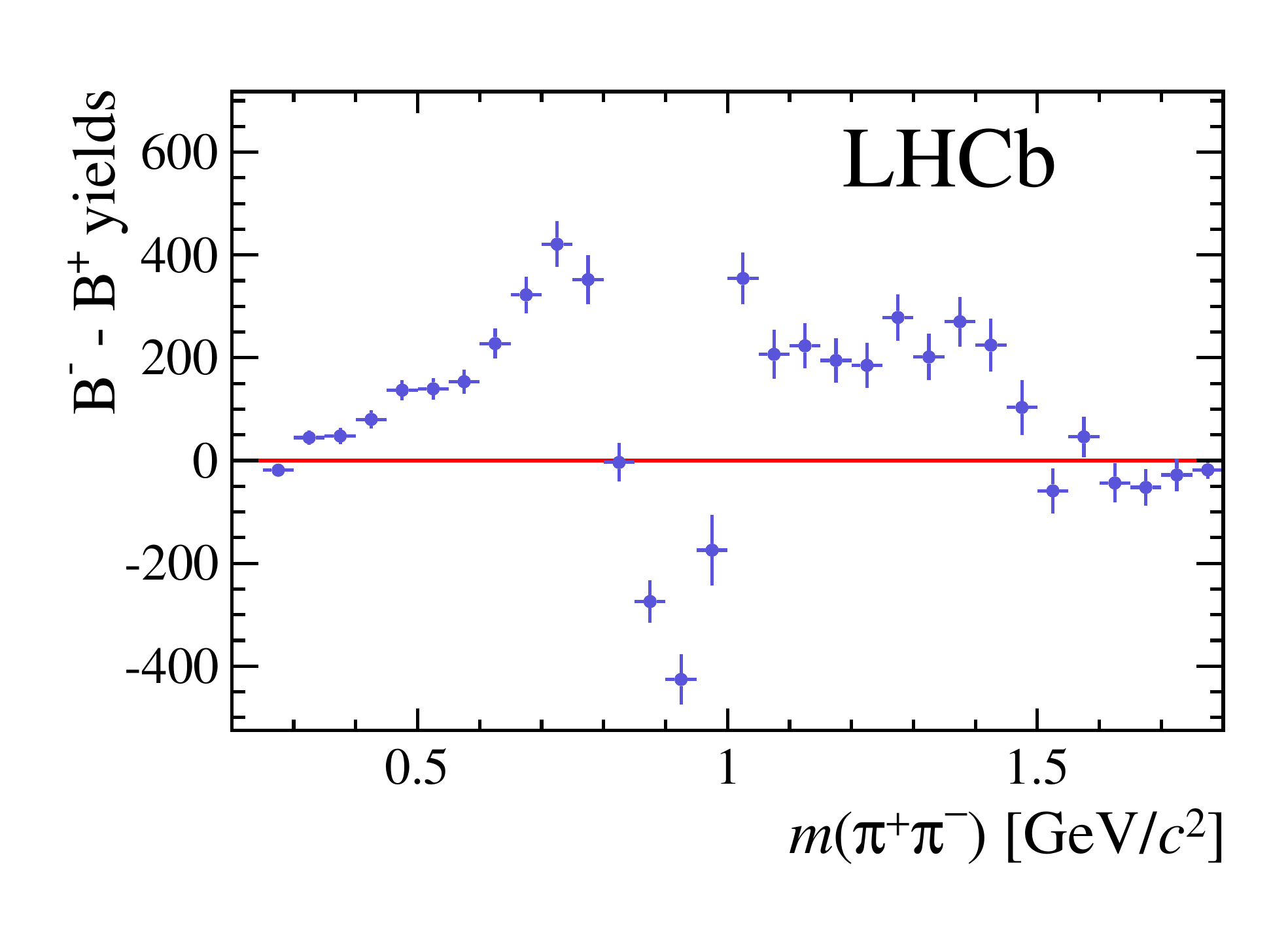}
 \put(85,58){\bf{(c)}}
\end{overpic}
\begin{overpic}[width=0.48\linewidth]{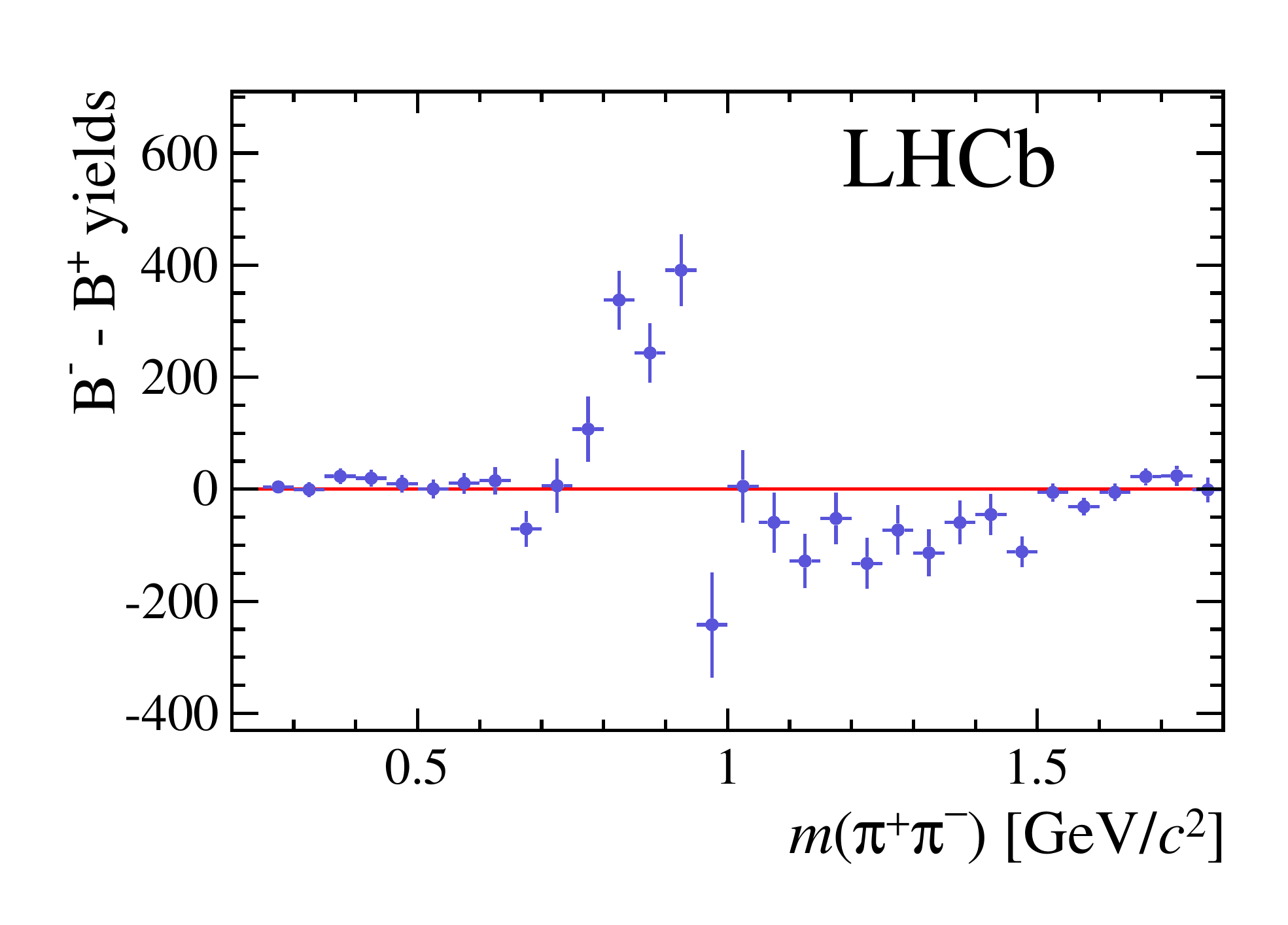}
 \put(85,58){\bf{(d)}}
\end{overpic}
\caption{
Projections in bins of the \mpipi variable of (a, b) the  number of \Bm and \Bp signal events and (c, d) their difference for \kpipi decays.
The plots are restricted to events with (a, c) $\cos\theta<0$  and (b, d) $\cos\theta>0$. The yields are acceptance-corrected and background-subtracted.
A guide line for zero (horizontal red line) was included on plots (c, d).
}
\label{fig:massProjkpipi:rho}
\end{center}
\end{figure}

\begin{figure} [tb]
\begin{center}
\begin{overpic}[width=0.48\linewidth]{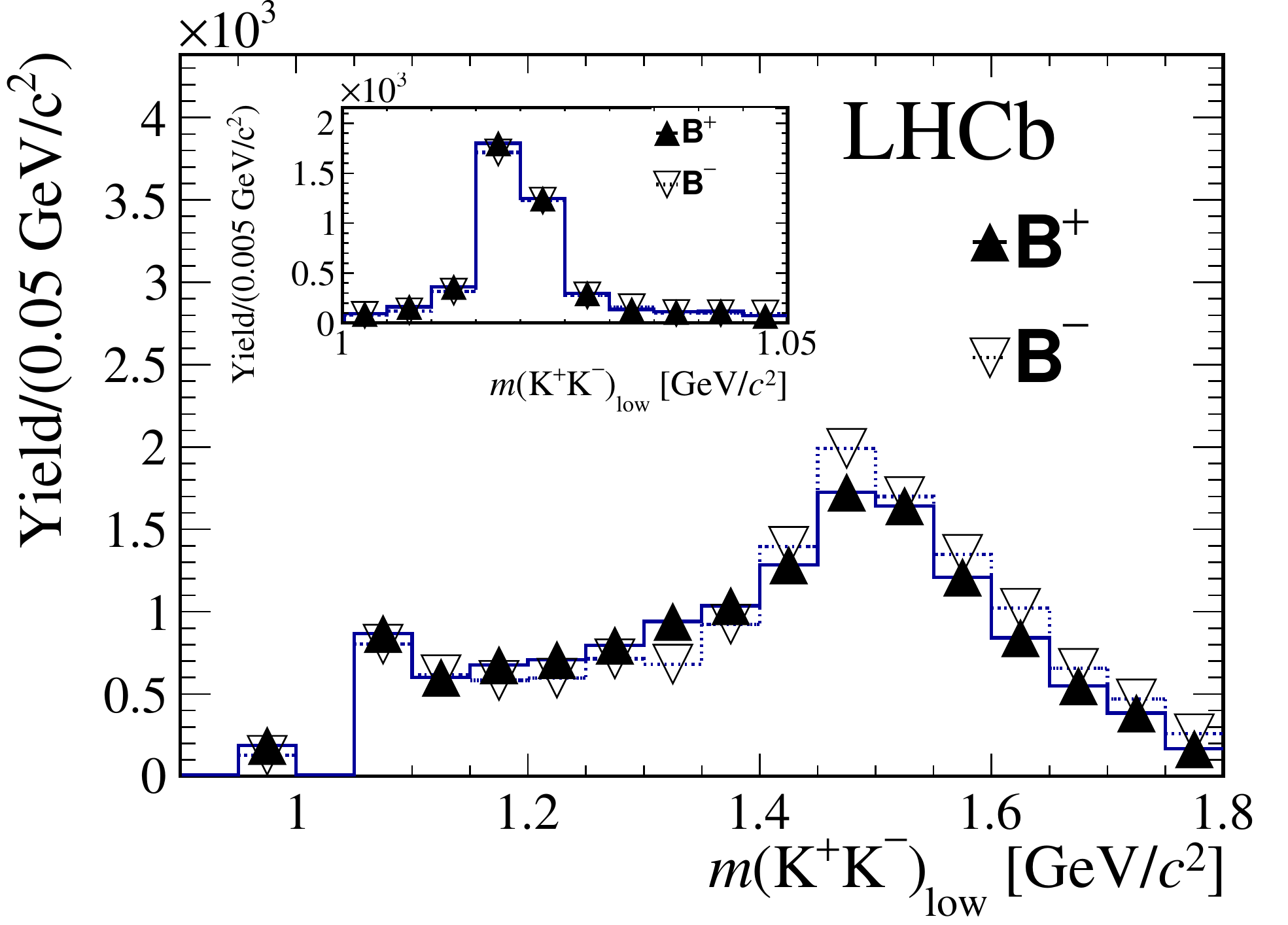}
\put(85,58){\bf{(a)}}
\end{overpic}
\begin{overpic}[width=0.48\linewidth]{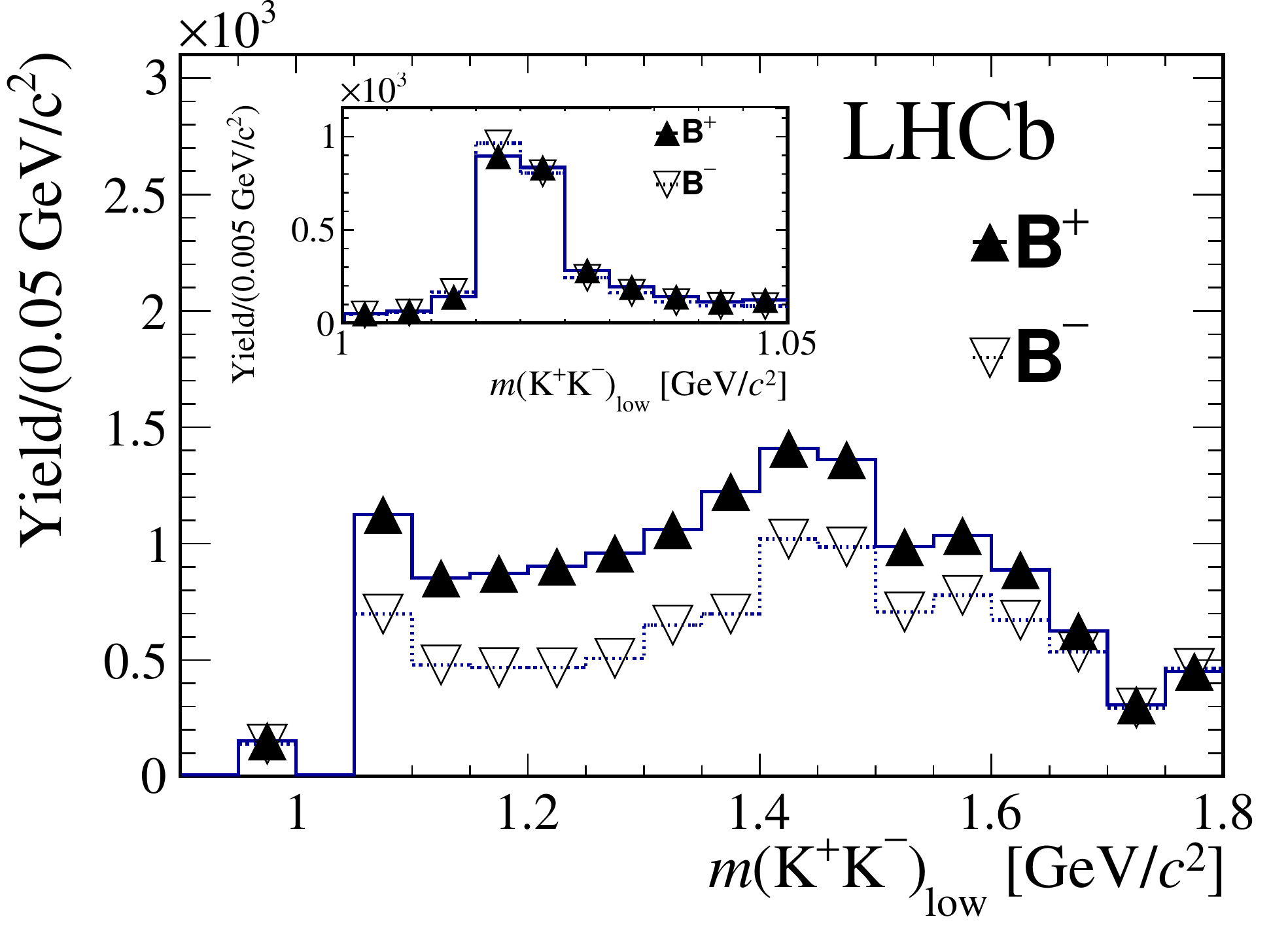}
 \put(85,58){\bf{(b)}}
\end{overpic}
\begin{overpic}[width=0.48\linewidth]{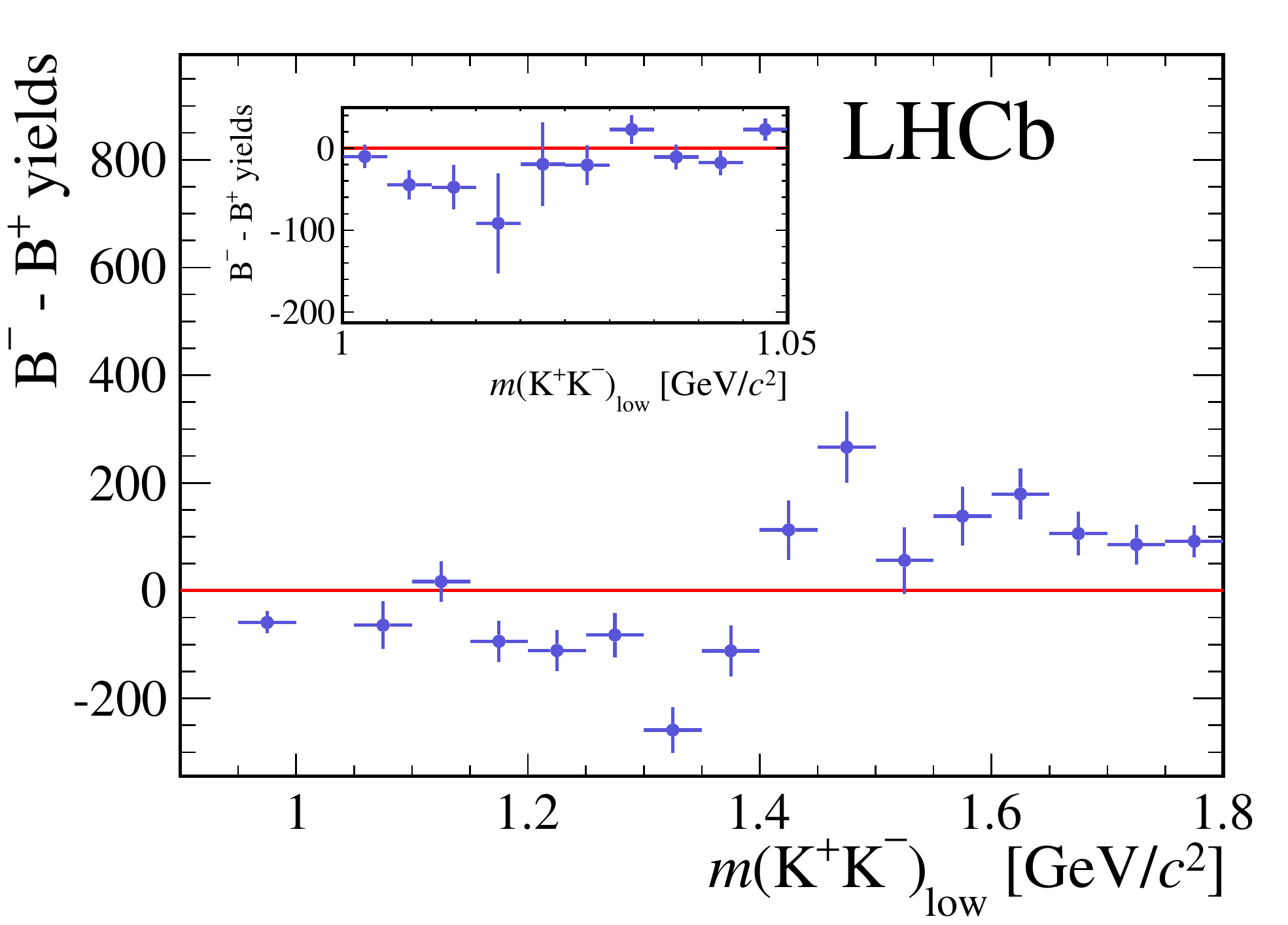}
 \put(85,58){\bf{(c)}}
\end{overpic}
\begin{overpic}[width=0.48\linewidth]{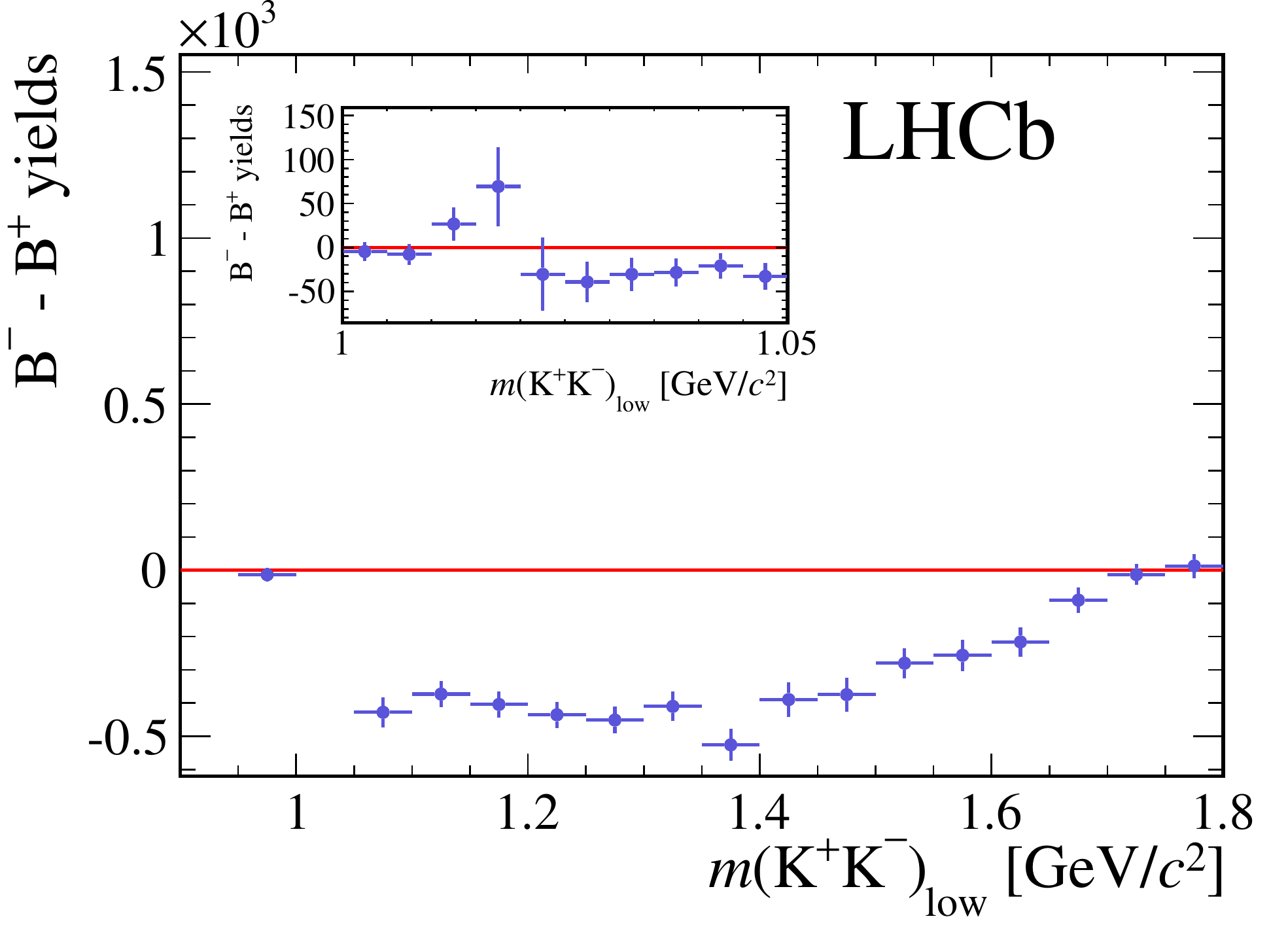}
 \put(85,58){\bf{(d)}}
\end{overpic}
\caption{
Projections in bins of the \mkklow variable of (a, b) the  number of \Bm and \Bp signal events and (c, d) their difference for \kkk decays. The inset plots show the $\phi$ resonance region of \mkklow between 1.00 and $1.05\gevcc$, which is excluded from the main plots.
The plots are restricted to events with (a, c) $\cos\theta<0$  and (b, d) $\cos\theta>0$.
The yields are acceptance-corrected and background-subtracted.
A guide line for zero (horizontal red line) was included on plots (c, d).
}
\label{fig:massProjkkk:phi2}
\end{center}
\end{figure}

\begin{figure}[tb]
\begin{center}
\begin{overpic}[width=0.48\linewidth]{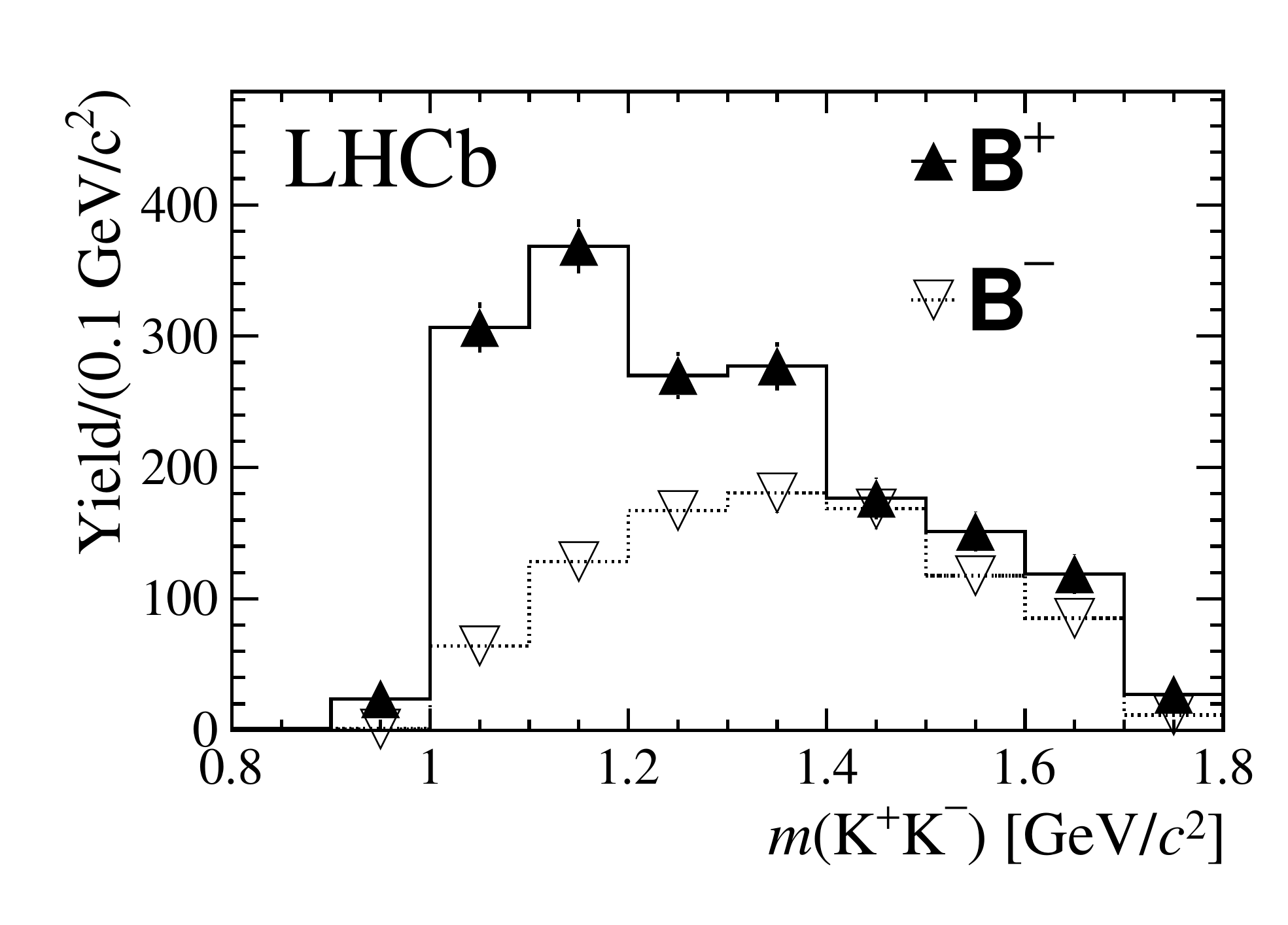}
\put(85,58){\bf{(a)}}
\end{overpic}
\begin{overpic}[width=0.48\linewidth]{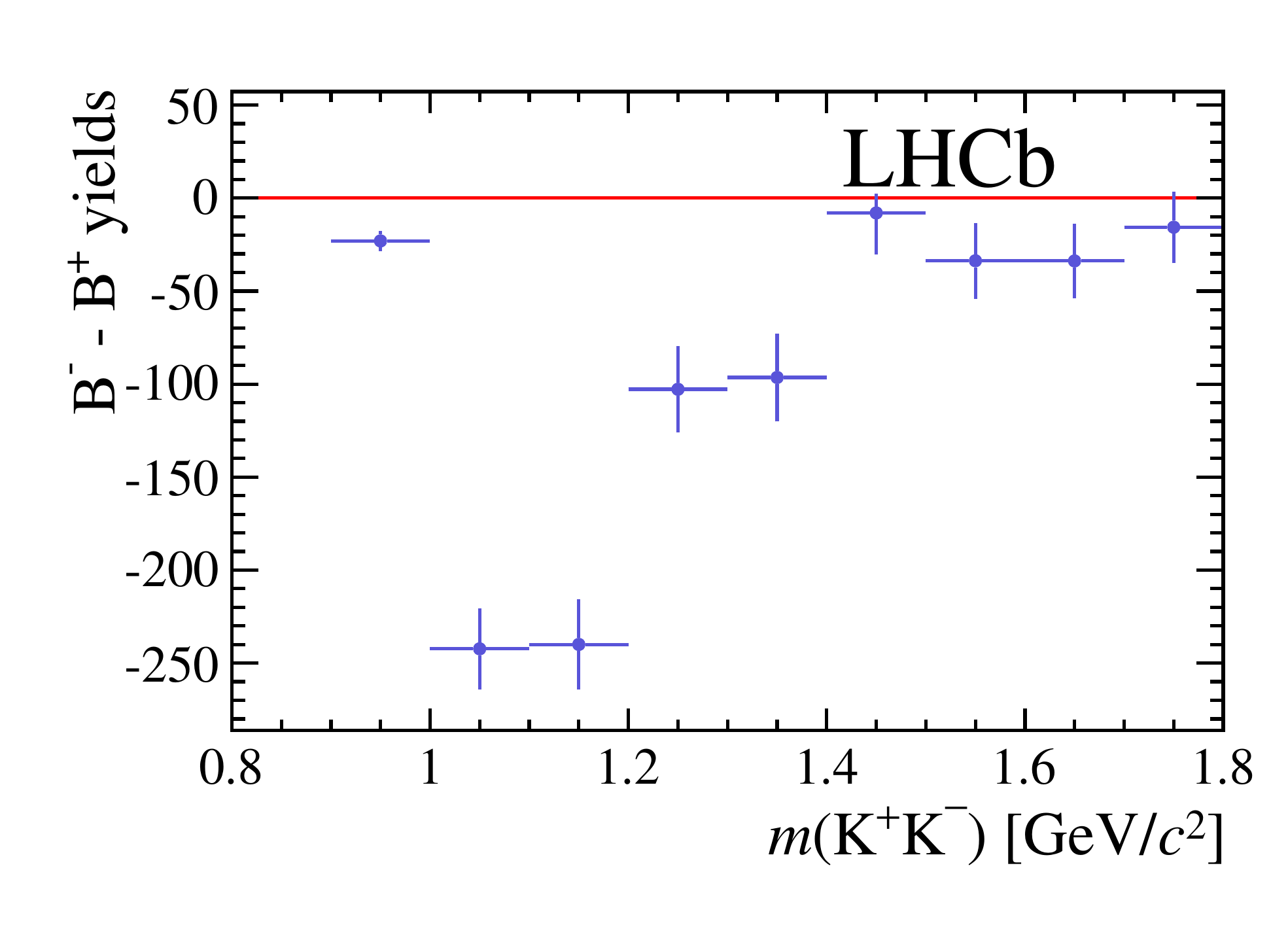}
\put(85,58){\bf{(b)}}
\end{overpic}

\caption{
Projections in bins of the \mkk variable of (a) the  number of \Bm and \Bp signal events and (b) their difference for \kkpi decays. The yields are acceptance-corrected and background-subtracted.
A guide line for zero (horizontal red line) was included on plot (b).
}
\label{fig:massProjkkpi:struct}
\end{center}
\end{figure}

The \acpn distributions in the Dalitz plots reveal rich structures, which are more evident in the two-body invariant-mass projection plots.
These are shown in Figs.~\ref{fig:massProjpipipi:rho} and~\ref{fig:massProjkpipi:rho} for the region of the $\rho$ resonance in \pipipi and \kpipi decays, respectively. The projections are split according to the sign of $\cos\theta$, where $\theta$ is the angle between the momenta of the unpaired hadron and the resonance decay product with the same-sign charge.
Figure~\ref{fig:massProjkkk:phi2} shows the projection onto the low $\Kp\Km$ invariant mass for the \kkk channel, while   Fig.~\ref{fig:massProjkkpi:struct} shows the projection into \mkk for the \kkpi mode.

The dynamic origin of the \CP asymmetries seen in Fig.~\ref{fig:AdaptiveACP}  can only be  fully understood with an amplitude analysis of these channels.
Nevertheless, the projections presented in  Figs.~\ref{fig:massProjpipipi:rho},~\ref{fig:massProjkpipi:rho},~\ref{fig:massProjkkk:phi2} and~\ref{fig:massProjkkpi:struct} indicate two different sources of \CP violation.
The first one may be associated with the $\pi^+\pi^- \leftrightarrow K^+K^- $ rescattering strong-phase difference in the region around $1.0$ to $1.5 \gevcc$~\cite{LHCb-PAPER-2013-027,LHCb-PAPER-2013-051}. In this region, there are more \Bm than \Bp decays into final states including a $\pi^+\pi^-$ pair (positive \CP asymmetry) and more $B^+$ than $B^-$ into final states that include a $K^+K^-$ pair (negative \CP asymmetry).
The second source of \CP violation, observed in both \kpipi and \pipipi decays around the $\rho(770)$ mass region, can be attributed to the final-state interference between the S-wave and P-wave in the Dalitz plot.

\subsection{\boldmath{\CP} asymmetry induced by rescattering}

Previous publications~\cite{LHCb-PAPER-2013-027,LHCb-PAPER-2013-051} showed evidence for  possible source of \CP violation produced by the long-distance strong phase through   $\pi^+\pi^- \leftrightarrow K^+K^-$ rescattering.
This interaction plays an important role in S-wave $\pi^+\pi^-$ elastic scattering, as was observed by previous experiments~\cite{PhysRevD.22.2595,Martin1979520} in the $\mpipi$  mass region between $1.0$ and $1.5 \gevcc$.
The \CPT symmetry requires that the sum of partial widths of a family of final states related to each other by strong rescattering, such as the four channels analysed here, are identical for particle and antiparticle.
As a consequence, positive \CP asymmetry in some channels implies negative \CP asymmetry in other channels of the same family.

\begin{figure}[tb]
\centering
\begin{overpic}[width=0.49\linewidth]{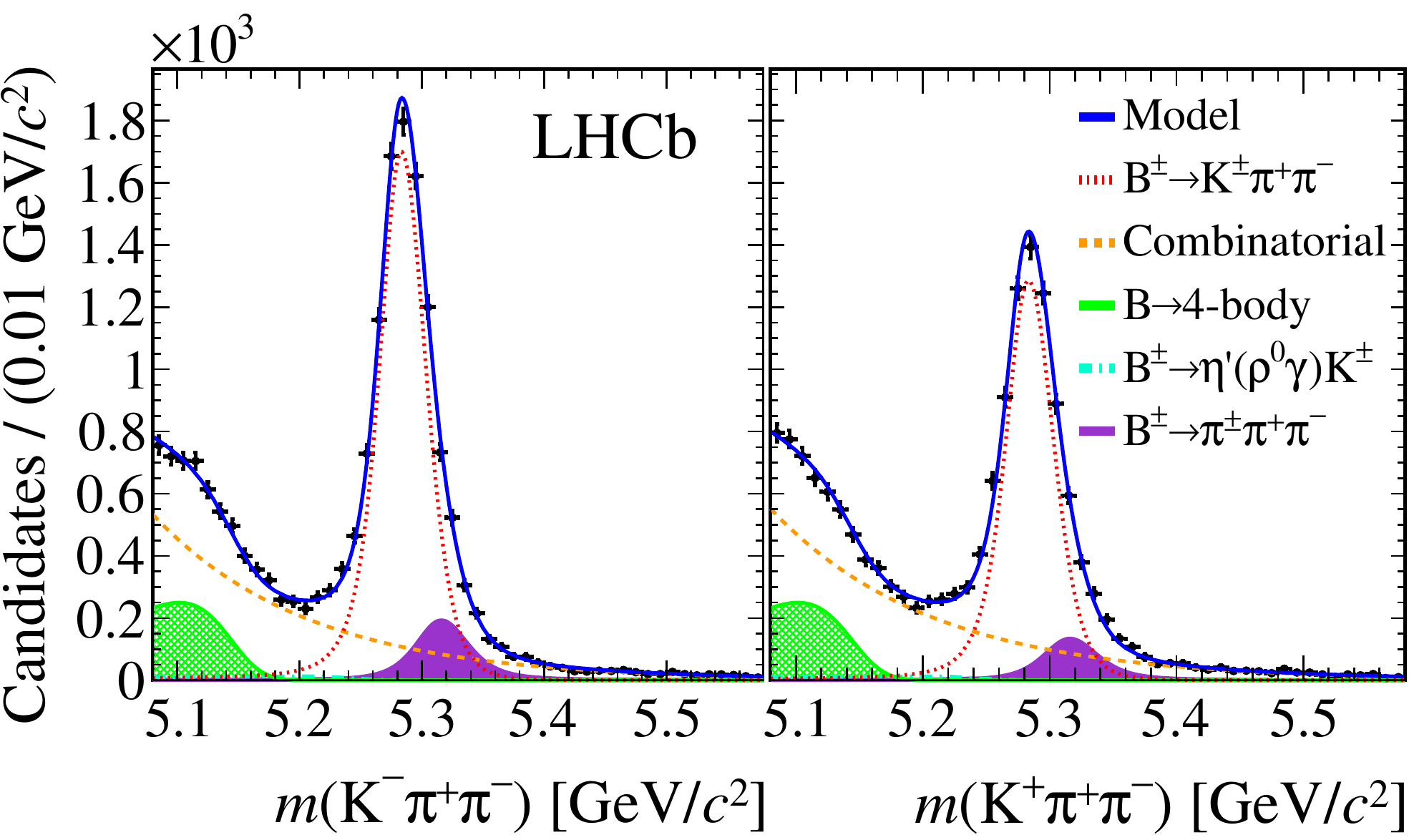}
\put(15,45){\bf{(a)}}
\end{overpic}
\begin{overpic}[width=0.49\linewidth]{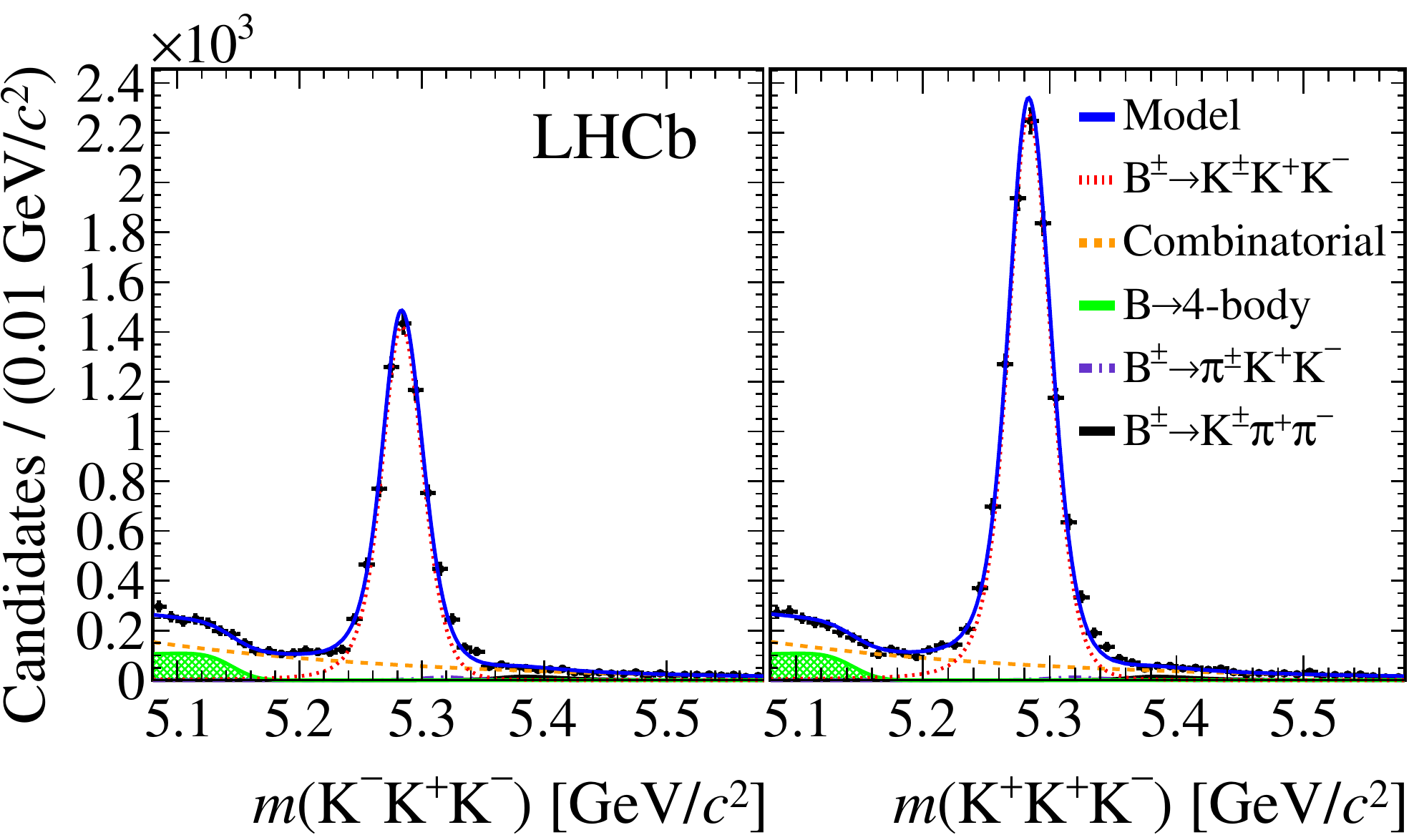}
\put(15,45){\bf{(b)}}
\end{overpic}
\begin{overpic}[width=0.49\linewidth]{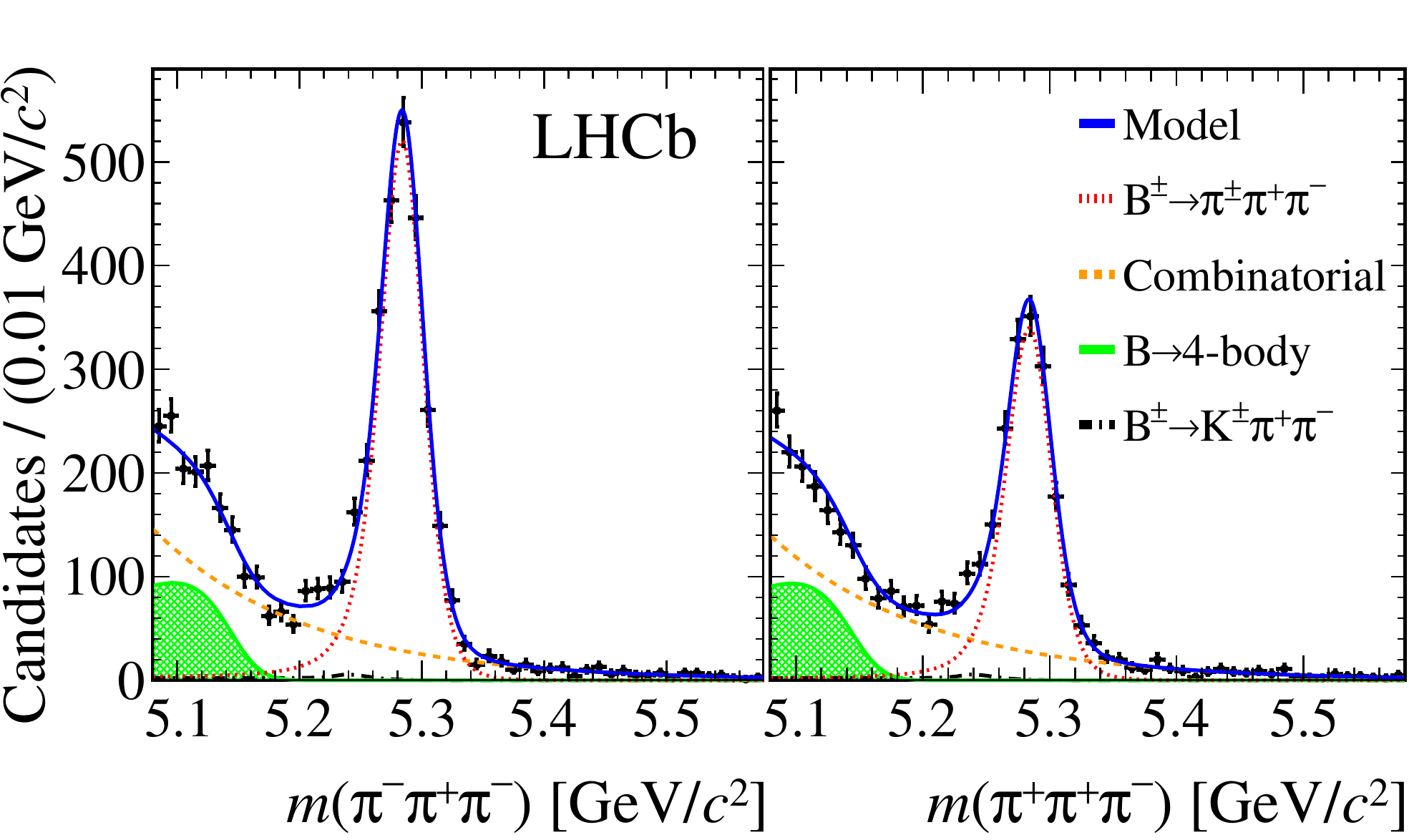}
\put(15,45){\bf{(c)}}
\end{overpic}
\begin{overpic}[width=0.49\linewidth]{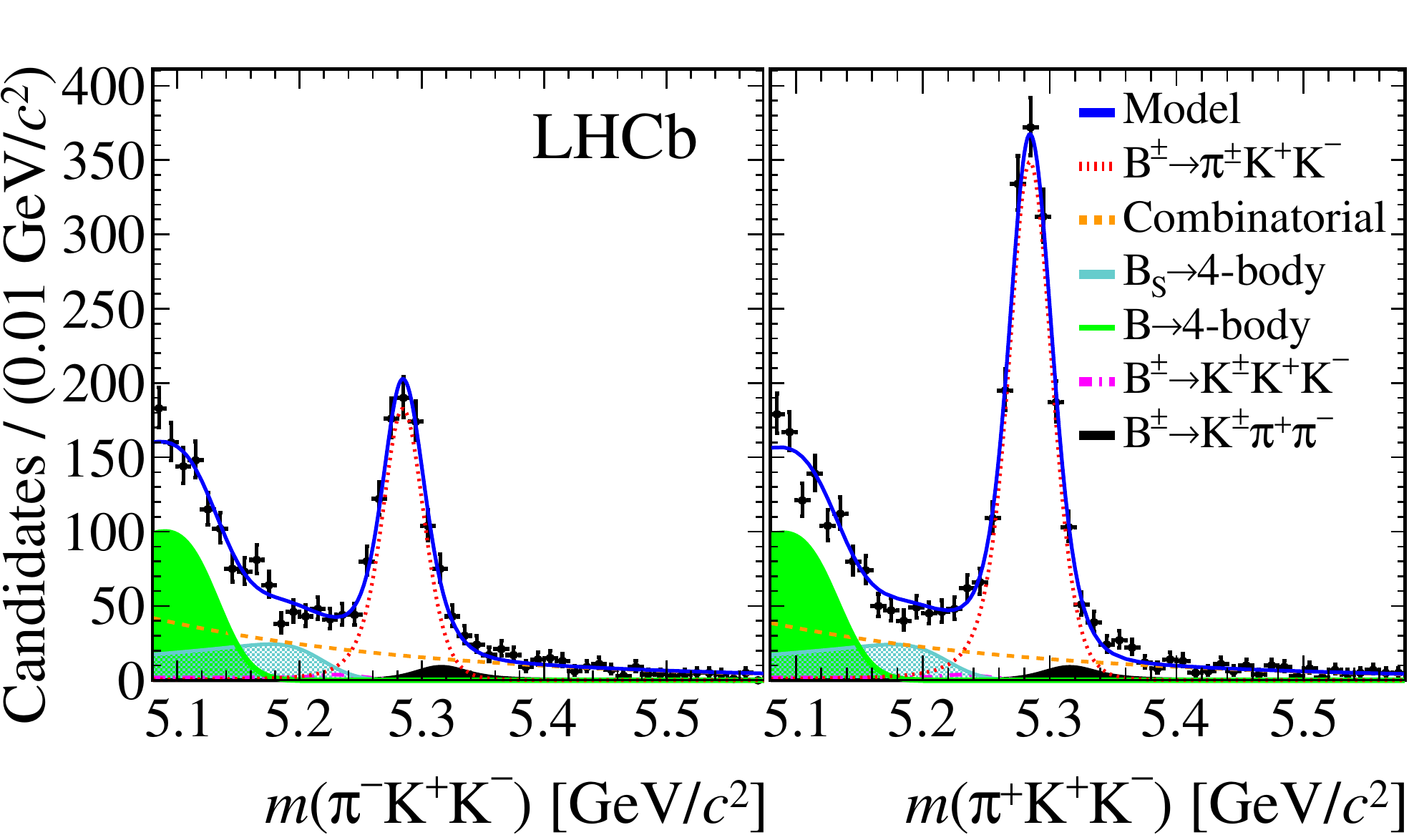}
\put(15,45){\bf{(d)}}
\end{overpic}
\caption{Invariant mass distributions in the rescattering regions ($\mpipi$ or $\mkk$ between $1.0$ and $1.5\gevcc$) for (a) \kpipi, (b) \kkk, (c) \pipipi and (d)~\kkpi decays.
The left panel in each figure shows the $B^{-}$ candidates and the right panel shows the $B^{+}$ candidates.
} \label{fig:s_phaseSpace:scat-combined2011and2012}
\end{figure}

The large data samples in the present study allow this effect to become evident, as shown in Figs.~\ref{fig:massProjpipipi:rho},~\ref{fig:massProjkpipi:rho},~\ref{fig:massProjkkk:phi2} and~\ref{fig:massProjkkpi:struct}.
Large asymmetries are observed for all the final states in the region between $1.0$ and $1.5 \gevcc$.
Figure~\ref{fig:s_phaseSpace:scat-combined2011and2012} shows the invariant mass distributions for events with \mpipi and \mkk in this interval, excluding the $\phi$ meson mass region for the \kkk mode.
The measured \CP asymmetries corresponding to the figure are given in Table~\ref{tab:scat-ACP-final}.
Decays involving a  $K^+K^-$ pair in the final state have a larger \CP asymmetry than their partner channels with a $\pip\pim$ pair.
The asymmetries are positive for channels with a $\pip\pim$ pair and negative for those with a $\Kp\Km$ pair.
This indicates that the mechanism of $\pi^+\pi^- \leftrightarrow K^+K^-$ rescattering could play an important role for \CP violation in charmless three-body \Bpm decays.

\begingroup
\begin{table}[tb]
\caption{Signal yields and charge asymmetries in the rescattering regions of \mpipi or \mkk  between $1.0$ and $1.5 \gevcc$. 
For the charge asymmetries, the first uncertainty is statistical, the second systematic, and the third is due to the \CP asymmetry of the \jpsik reference mode.}
\begin{center}
\begin{tabular}{lcc}
\hline
 Decay & $N_S$ &  $A_{\CP}$ \\ \midrule
\kpipi& $15\,562 \pm 165$ & $+0.121 \pm 0.012 \pm 0.017 \pm 0.007$ \\
\kkk& $16\,992 \pm 142$ & $-0.211 \pm 0.011 \pm 0.004 \pm 0.007$ \\
\pipipi& $\,4329 \pm 76$ & $+0.172 \pm 0.021 \pm 0.015 \pm 0.007$ \\
\kkpi& $\,2500 \pm 57$ & $-0.328 \pm 0.028 \pm 0.029 \pm 0.007$ \\
\hline
\end{tabular} \end{center}
\label{tab:scat-ACP-final}\end{table}
\endgroup

\subsection{\boldmath{\CP} asymmetry due to interference between partial waves}

In hadronic three-body decays, there is another long-distance strong-interaction phase, which stems from the amplitudes of  intermediate resonances.
The Breit-Wigner propagator associated with an intermediate resonance can provide a phase that varies with the resonance mass.
There is also a phase related to final-state interactions, associated with each intermediate state that contributes to the same final state.
In general, the latter phase is considered constant within the phase space.
This phase also includes any short-distance strong phase.

These three sources of strong phases are known to give clear signatures in the Dalitz plane~\cite{Miranda1, BGM}.
The short-distance direct \CP violation is proportional to the difference of the magnitude between the positive and negative amplitudes of the resonance, and is therefore proportional to the square of the Breit-Wigner propagator associated with the resonance.

The interference term has two components.
One  is associated with the real part of the Breit-Wigner propagator and is directly proportional to $(m^2_R - s)$, where $m_R$ is the central value of the resonance mass and $s$ is the square of the invariant mass of its decay products.
The other component is proportional to the product $m_R \Gamma $, where $\Gamma$ is the width of the resonance.
The relative proportion of real and imaginary terms of these interference components gives the final-state interaction phase difference between the two amplitudes.

Another  feature of the interference term is the characteristic angular distribution.
For a decay involving one vector resonance, the interference term is multiplied by $\cos\theta$, which is a linear function of the other Dalitz variable.
As the cosine varies from $-1$ to 1, the interference term changes sign around the middle of the Dalitz plot.
The short-distance \CP violation does not change sign because it is proportional to the  square of the amplitude ($\cos^2\theta$).

The charge asymmetry in \pipipi decays changes sign, as shown in Fig.~\ref{fig:massProjpipipi:rho}, at a value of \mpipi close to the $\rho(770)$ resonance.
This is an indication of the dominance of the  long-distance interference effect in this region of the Dalitz plot.
Moreover, since this change of sign occurs for both $\cos\theta > 0$ and $\cos\theta < 0$,  the dominant term of the Dalitz interference is inferred to be the real part of the Breit-Wigner propagator.

When the interference between P-wave and S-wave involves two resonances, as for \kpipi decays, with $\rho(770)$ and  $f_0(980)$ resonances, the dominant component of the real Dalitz \CP asymmetry is proportional to a product of the type $(m^2_{\rho} -s )(m^2_{f_0} - s)$, thus having two zeros, as can be seen in Fig.~\ref{fig:massProjkpipi:rho}.
In the $\cos\theta < 0$ region there is a zero around the $\rho(770)$ mass and another one around the $f_0(980)$ meson mass. However, in the region of $\cos\theta > 0$, a clear change of sign is only seen around the $f_0(980)$ mass.

The yield of the $f_0(980)$ resonance in \kpipi decays depends on $\cos\theta$ (Fig.~\ref{fig:massProjkpipi:rho}).
The yield around the $f_0(980)$ mass for $\cos\theta > 0$ is almost twice that of the region with $\cos\theta < 0$.
The magnitude of the \CP asymmetry indicates the opposite dependence.
Also, the yield around the $\rho(770)$ resonance in the \pipipi decay changes significantly in the two $\cos\theta$ regions (Fig.~\ref{fig:massProjpipipi:rho}).
The largest yield in the $\rho(770)$ mass region with a small \CP asymmetry occurs for $\cos\theta < 0$, while  there is a large \CP asymmetry with fewer events in the $\rho(770)$ mass region for $\cos\theta > 0$.
One possible explanation is that the fractions of tree and penguin contributions may vary across the phase space~\cite{PhysRevD.88.114014}.
Understanding this effect may be important to performing amplitude analyses.

\begin{figure}[tb]
\centering
\begin{overpic}[width=0.49\linewidth]{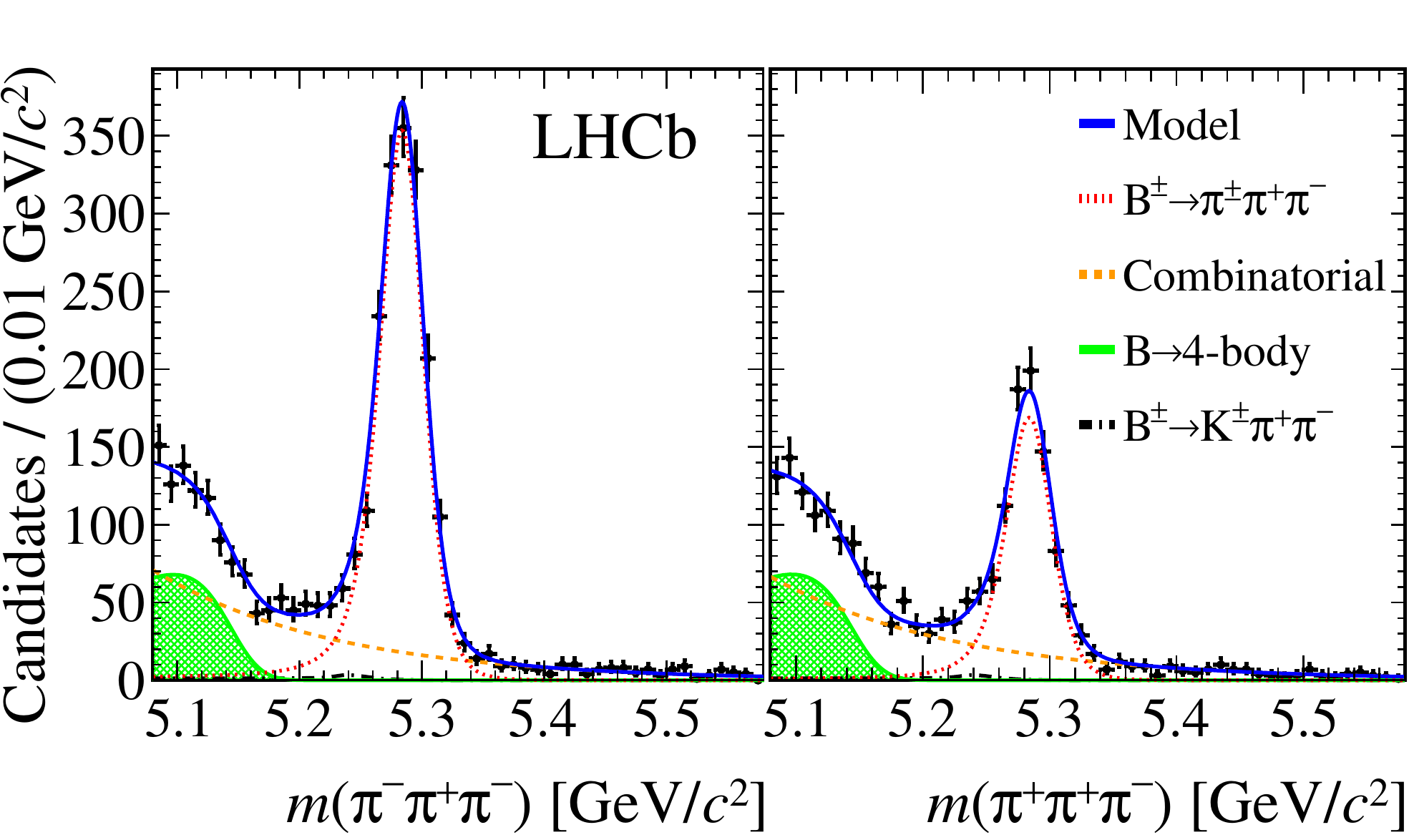}
\put(15,45){\bf{(a)}}
\end{overpic}
\begin{overpic}[width=0.49\linewidth]{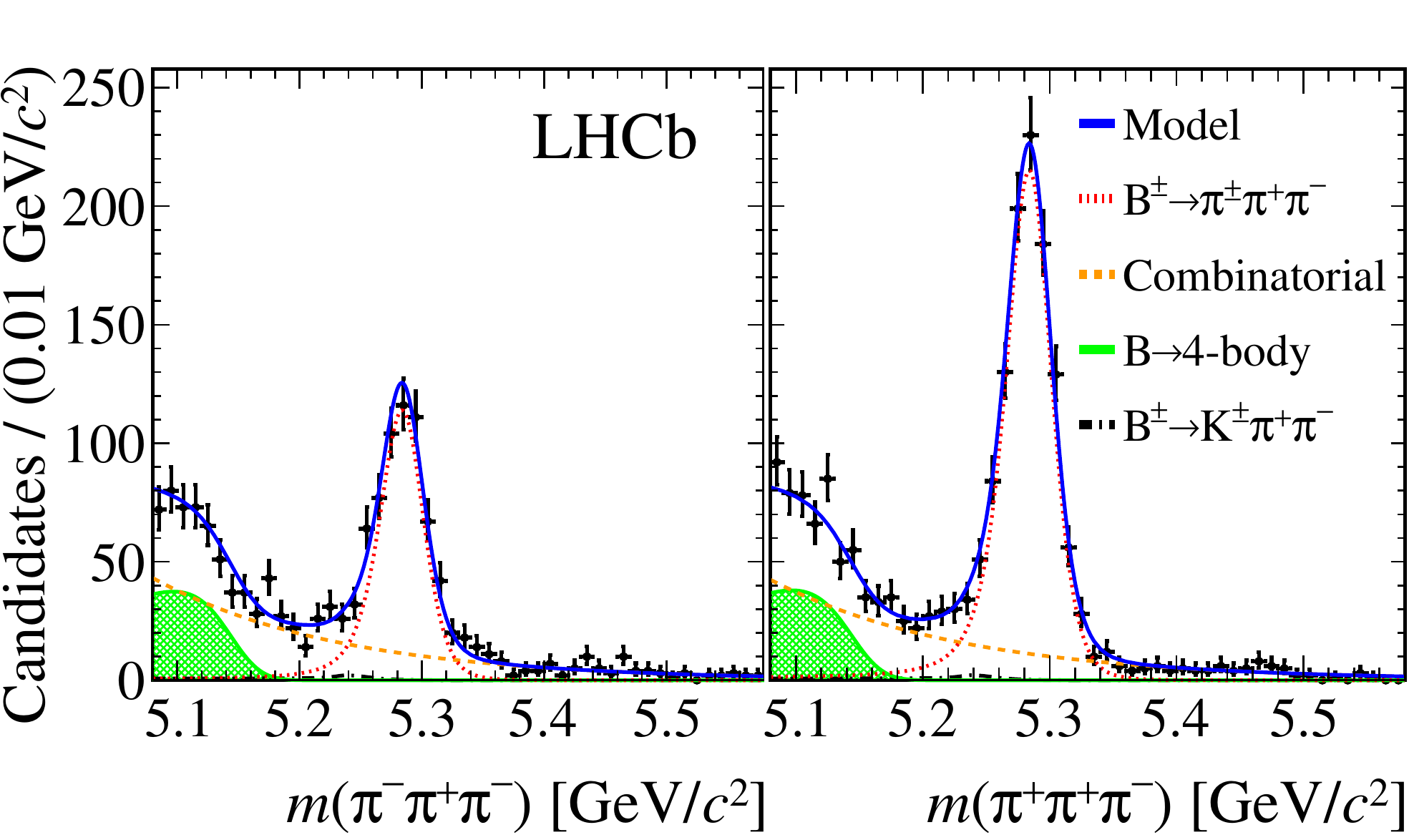}
\put(15,45){\bf{(b)}}
\end{overpic}
\begin{overpic}[width=0.49\linewidth]{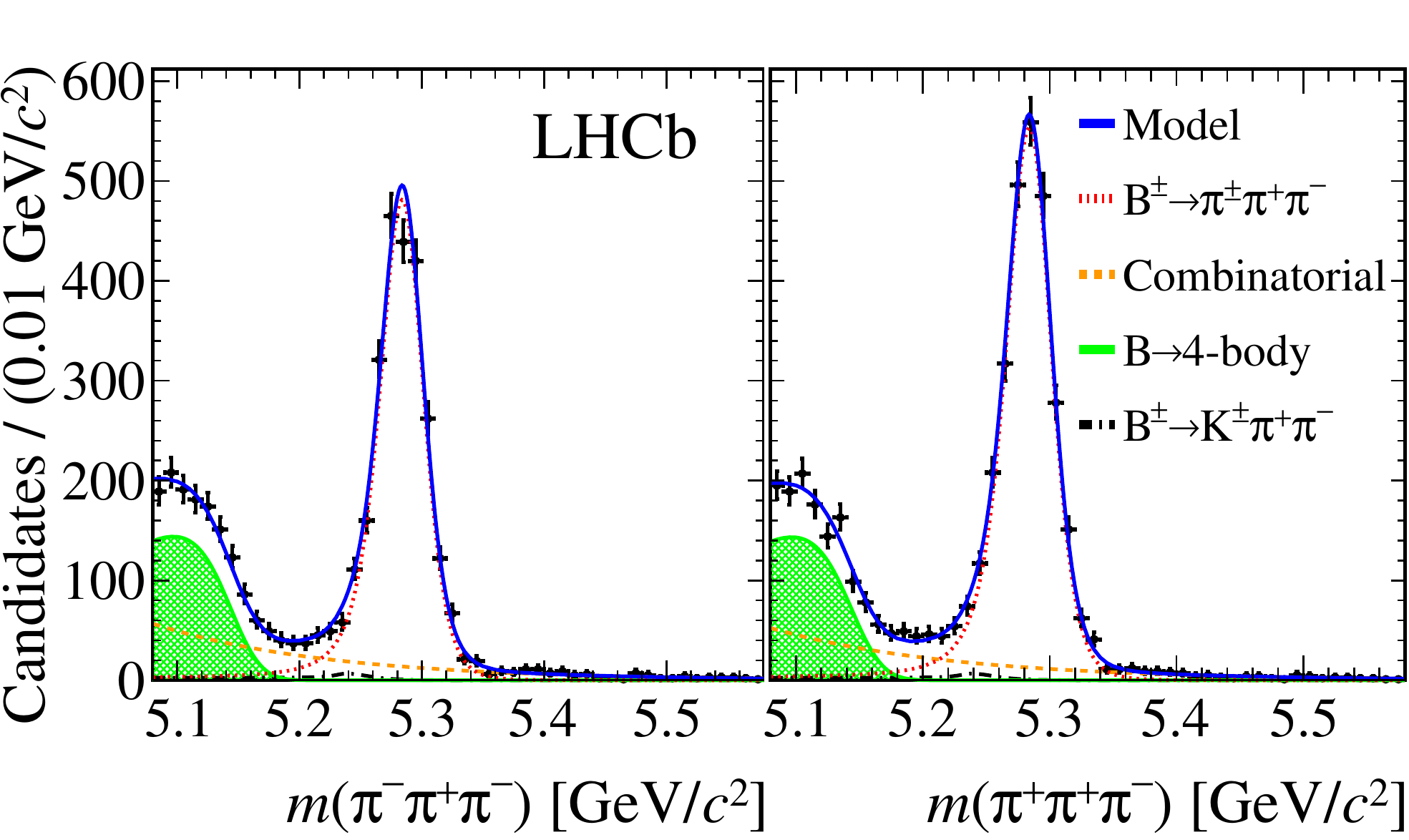}
\put(15,45){\bf{(c)}}
\end{overpic}
\begin{overpic}[width=0.49\linewidth]{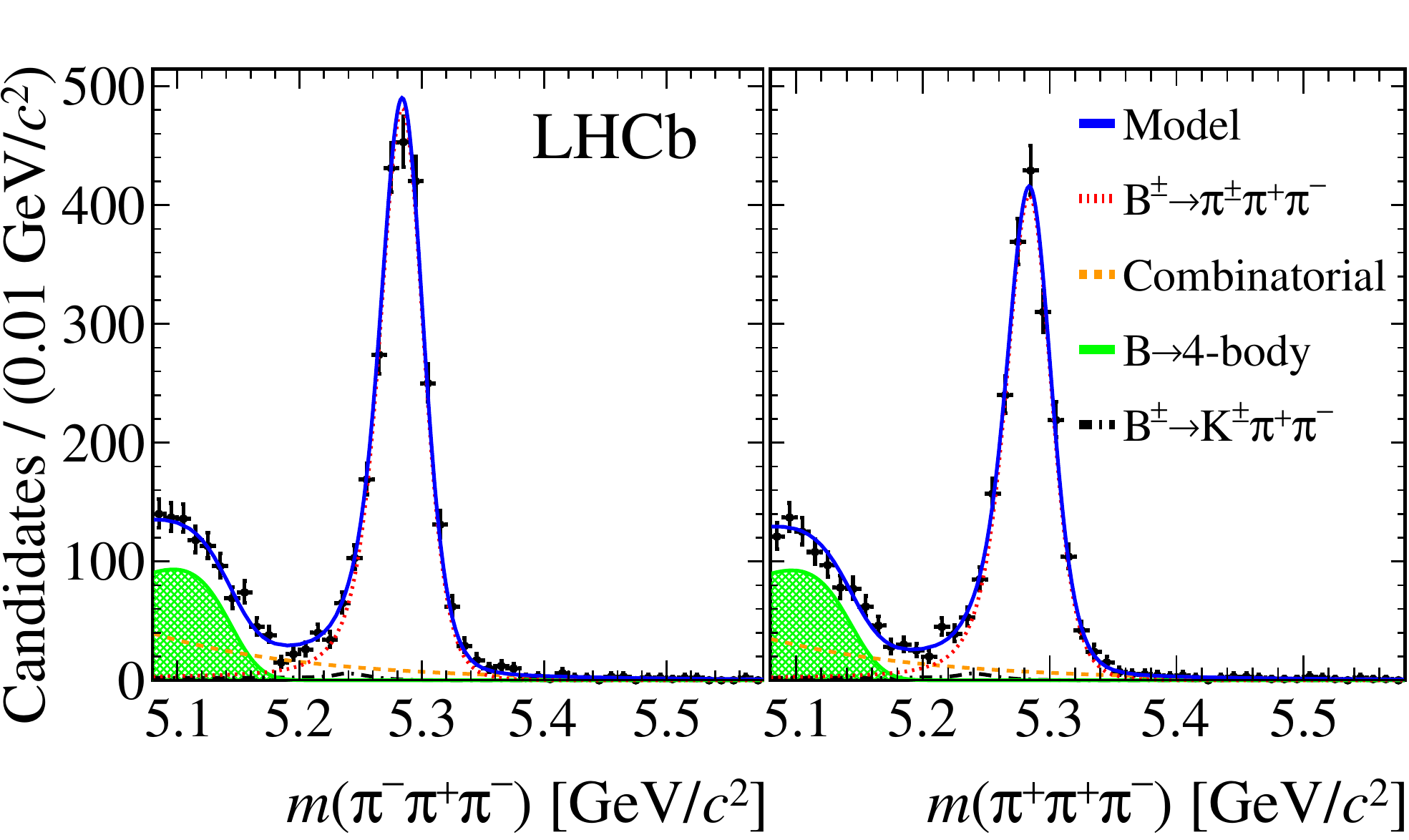}
\put(15,45){\bf{(d)}}
\end{overpic}
\caption{Invariant mass distribution of \pipipi candidates restricted to (a) sector I, (b) sector II, (c) sector III and (d) sector IV. The left panel in each figure shows the $B^{-}$ candidates and the right panel shows the $B^{+}$ candidates.
}
\label{fig:s_phaseSpace:rho-combined2011and2012}
\end{figure}

\begin{table*}[tb]
\caption{Signal yields and charge asymmetries in the regions dominated by the vector resonances.
For the charge asymmetries, the first uncertainty is statistical, the second systematic, and the third is due to the \CP asymmetry of the \jpsik reference mode.}
\resizebox{\linewidth}{!}{
\begin{tabular}{lcccc}
\hline
 Decay mode & Resonance & Sector & $N_S$ &$A_{\CP}$ \\ \midrule
\multirow{4}{*}{\kpipi}  & \multirow{4}{*}{$\rho$} &  I & $2909 \pm 80$ & $-0.052 \pm 0.032 \pm 0.047 \pm 0.007$ \\
& &  II & $\;\;6136 \pm 100$ & $+0.140 \pm 0.018 \pm 0.034 \pm 0.007$ \\
& &  III & $2856 \pm 86$ & $+0.598 \pm 0.036 \pm 0.079 \pm 0.007$ \\
& & IV& $2107 \pm 55$ & $-0.208 \pm 0.043 \pm 0.042 \pm 0.007$ \\
\midrule
\multirow{4}{*}{\kpipi}  & \multirow{4}{*}{\Kstar} &  I & $11\,095 \pm 115$ & $+0.002 \pm 0.013 \pm 0.011 \pm 0.007$ \\
& &  II & $7159 \pm 89$ & $+0.007 \pm 0.016 \pm 0.005 \pm 0.007$ \\
& &  III & $2427 \pm 65$ & $-0.009 \pm 0.031 \pm 0.054 \pm 0.007$ \\
& & IV& $\;\;9861 \pm 124$ & $-0.020 \pm 0.015 \pm 0.010 \pm 0.007$ \\
\midrule
\multirow{4}{*}{\pipipi}  & \multirow{4}{*}{$\rho$} &  I & $2629 \pm 59$ & $+0.302 \pm 0.026 \pm 0.015 \pm 0.007$ \\
& &  II & $1653 \pm 46$ & $-0.244 \pm 0.034 \pm 0.019 \pm 0.007$ \\
& &  III & $5204 \pm 79$ & $-0.076 \pm 0.019 \pm 0.007 \pm 0.007$ \\
& & IV& $4476 \pm 72$ & $+0.055 \pm 0.020 \pm 0.013 \pm 0.007$ \\
\midrule
\multirow{4}{*}{\kkk}  & \multirow{4}{*}{$\phi$} &  I & $3082 \pm 56$ & $-0.018 \pm 0.024 \pm 0.008 \pm 0.007$ \\
& &  II & $4119 \pm 64$ & $-0.008 \pm 0.021 \pm 0.004 \pm 0.007$ \\
& &  III & $1546 \pm 39$ & $+0.066 \pm 0.034 \pm 0.010 \pm 0.007$ \\
& & IV& $2719 \pm 53$ & $+0.015 \pm 0.026 \pm 0.002 \pm 0.007$ \\
\hline
\end{tabular}
}
\label{tab:reg-ACP-final}\end{table*}

The \CP asymmetry  around the $\rho(770)$ peak in $ \pi^+\pi^-$ invariant mass changes sign depending on the invariant mass and angular distribution.
To quantify the \CP violation in the region where vector particles contribute without losing sensitivity to the interference asymmetry behaviour, the region is divided into four sectors.
Sectors I and III are on the low-mass side of the resonance mass ($0.47 < \mpipilow < 0.77\gevcc$ for \pipipi decays), while sectors II and IV are on the high-mass side ($0.77 < \mpipilow < 0.92\gevcc$).
Sectors I and II are delimited by $\cos\theta > 0$ (upper part of the Dalitz plot), while sectors III and IV are delimited by $\cos\theta < 0 $ (lower part).

Figure~\ref{fig:s_phaseSpace:rho-combined2011and2012} shows the invariant mass distributions for \pipipi decays divided into these four sectors.
The charge asymmetries are measured using the same method as for the inclusive asymmetries.
Similar measurements of the \CP asymmetries are performed for events restricted to regions dominated by the $\rho(770)$ and $K^*(892)$ resonances in  \kpipi decays, and the $\phi(1020)$ resonance in \kkk decays, with the results given in Table~\ref{tab:reg-ACP-final}.
Only the decays involving the $\rho(770)$ resonance have a significant \CP asymmetry.
The $K^*(892)$ charge asymmetry is consistent with zero, as expected and for the $\phi(1020)$ resonance, the results are in agreement with the previous \lhcb analysis~\cite{LHCb-PAPER-2013-048}.

\section{Conclusion}

We measure the inclusive \CP asymmetries for the four charmless three-body charged decays \kpipi, \kkk, \pipipi and \kkpi,
\begin{eqnarray}
\acp(\kpipi)&=& +\kpipiacp \, , \nonumber \\
\acp(\kkk) &=& \kkkacp \, , \nonumber \\
\acp(\pipipi)&=& +\pipipiacp \, , \nonumber \\
\acp(\kkpi)&=& \kkpiacp \, , \nonumber
\end{eqnarray}
where the first uncertainty is statistical, the second systematic, and the third is due to the \CP asymmetry of the \jpsik reference mode, with significances of $\kpipisigma\sigma$, $\kkksigma\sigma$, $\pipipisigma\sigma$ and $\kkpisigma\sigma$, respectively.
The results, which are obtained from an analysis of data corresponding to an integrated luminosity of 3.0\invfb,  are consistent with and supersede  the previous \lhcb analyses based on 1.0\invfb of data~\cite{LHCb-PAPER-2013-027,LHCb-PAPER-2013-051}.

The \CP asymmetries are not uniformly distributed in the phase space.
For each of the channels, we observe a significant \CP asymmetry in the $\mkk$ or $\mpipi$ invariant mass region between $1.0$ and $1.5 \gevcc$.
These \CP asymmetries are positive for the channels that include two pions in the final state and negative for those that include two kaons.
These results are in agreement with those from previous LHCb publications~\cite{LHCb-PAPER-2013-027,LHCb-PAPER-2013-051} and could be due to long-distance $\pi^+\pi^- \leftrightarrow \Kp\Km$ rescattering \cite{Bhattacharya:2013cvn, IgnacioCPT}.
A comprehensive study of this phenomenon has to involve the complete set of paired channels, including the $\Bpm \to \Kpm \pi^0\pi^0$ and $\Bpm \to \Kpm \Kzb \Kz$ decays for the \kpipi family, and the $\Bpm \to \pi^{\pm}\pi^0\pi^0$ and $\Bpm \to \pi^{\pm}\Kzb \Kz$ decays for the \pipipi one.
The role of the unpaired hadron in two-body $\pi^+\pi^- \leftrightarrow \Kp\Km$ rescattering merits further investigation~\cite{PhysRevD.84.094001}.

The \CP asymmetry related to the $\rho(770)$ resonance in the $\pi^+ \pi^-$ invariant mass below $1\gevcc$ is also reported.
The behaviour of this asymmetry, which crosses zero around the $\rho(770)$ mass in both \kpipi and \pipipi modes, indicates a  \CP asymmetry related to the real part of the long-distance interaction between the S-wave and P-wave contributions to $\pi^+ \pi^-$.
Further understanding of the resonance contributions and \CP asymmetries in these decays will require amplitude analyses.

\section*{Acknowledgements}

\noindent We express our gratitude to our colleagues in the CERN
accelerator departments for the excellent performance of the LHC. We
thank the technical and administrative staff at the LHCb
institutes. We acknowledge support from CERN and from the national
agencies: CAPES, CNPq, FAPERJ and FINEP (Brazil); NSFC (China);
CNRS/IN2P3 (France); BMBF, DFG, HGF and MPG (Germany); SFI (Ireland); INFN (Italy);
FOM and NWO (The Netherlands); MNiSW and NCN (Poland); MEN/IFA (Romania);
MinES and FANO (Russia); MinECo (Spain); SNSF and SER (Switzerland);
NASU (Ukraine); STFC (United Kingdom); NSF (USA).
The Tier1 computing centres are supported by IN2P3 (France), KIT and BMBF
(Germany), INFN (Italy), NWO and SURF (The Netherlands), PIC (Spain), GridPP
(United Kingdom).
We are indebted to the communities behind the multiple open
source software packages on which we depend. We are also thankful for the
computing resources and the access to software R\&D tools provided by Yandex LLC (Russia).
Individual groups or members have received support from
EPLANET, Marie Sk\l{}odowska-Curie Actions and ERC (European Union),
Conseil g\'{e}n\'{e}ral de Haute-Savoie, Labex ENIGMASS and OCEVU,
R\'{e}gion Auvergne (France), RFBR (Russia), XuntaGal and GENCAT (Spain), Royal Society and Royal
Commission for the Exhibition of 1851 (United Kingdom).

\addcontentsline{toc}{section}{References}
\setboolean{inbibliography}{true}
\bibliographystyle{LHCb}
\bibliography{main,LHCb-PAPER,LHCb-CONF,LHCb-DP,LHCb-TDR}

\newpage

\centerline{\large\bf LHCb collaboration}
\begin{flushleft}
\small
R.~Aaij$^{41}$, 
B.~Adeva$^{37}$, 
M.~Adinolfi$^{46}$, 
A.~Affolder$^{52}$, 
Z.~Ajaltouni$^{5}$, 
S.~Akar$^{6}$, 
J.~Albrecht$^{9}$, 
F.~Alessio$^{38}$, 
M.~Alexander$^{51}$, 
S.~Ali$^{41}$
% ...

{\footnotesize \it
% ...
$ ^{5}$Clermont Universit\'{e}, Universit\'{e} Blaise Pascal, CNRS/IN2P3, LPC, Clermont-Ferrand, France\\
$ ^{6}$CPPM, Aix-Marseille Universit\'{e}, CNRS/IN2P3, Marseille, France\\
% ...
$ ^{9}$Fakult\"{a}t Physik, Technische Universit\"{a}t Dortmund, Dortmund, Germany\\
% ...
$ ^{37}$Universidad de Santiago de Compostela, Santiago de Compostela, Spain\\
$ ^{38}$European Organization for Nuclear Research (CERN), Geneva, Switzerland\\
% ...
$ ^{41}$Nikhef National Institute for Subatomic Physics, Amsterdam, The Netherlands\\
% ...
$ ^{46}$H.H. Wills Physics Laboratory, University of Bristol, Bristol, United Kingdom\\
% ...
$ ^{51}$School of Physics and Astronomy, University of Glasgow, Glasgow, United Kingdom\\
$ ^{52}$Oliver Lodge Laboratory, University of Liverpool, Liverpool, United Kingdom\\
% ...
}
\end{flushleft}

\end{document}

%% file: lhcb-symbols-def.tex
\def\lhcb {\mbox{LHCb}\xspace}

\ifthenelse{\boolean{uprightparticles}}%
{

 \def\Pmu         {\ensuremath{\upmu}\xspace}

 \def\Ppi         {\ensuremath{\uppi}\xspace}

 \def\Ppsi        {\ensuremath{\uppsi}\xspace}

 \def\PDelta      {\ensuremath{\Delta}\xspace}                 
 \def\PXi      {\ensuremath{\Xi}\xspace}                 
 \def\PLambda      {\ensuremath{\Lambda}\xspace}                 
 \def\PSigma      {\ensuremath{\Sigma}\xspace}                 
 \def\POmega      {\ensuremath{\Omega}\xspace}                 
 \def\PUpsilon      {\ensuremath{\Upsilon}\xspace}                 
 
 \def\PB      {\ensuremath{\mathrm{B}}\xspace}                 
                  
 \def\PD      {\ensuremath{\mathrm{D}}\xspace}

 \def\PJ      {\ensuremath{\mathrm{J}}\xspace}                 
 \def\PK      {\ensuremath{\mathrm{K}}\xspace}

 \def\Pb      {\ensuremath{\mathrm{b}}\xspace}                 
 \def\Pc      {\ensuremath{\mathrm{c}}\xspace}

 \def\Pi      {\ensuremath{\mathrm{i}}\xspace}

}
{

 \def\Pmu         {\ensuremath{\mu}\xspace}

 \def\Ppi         {\ensuremath{\pi}\xspace}

 \def\Ppsi        {\ensuremath{\psi}\xspace}                 
                  
 \mathchardef\PDelta="7101
 \mathchardef\PXi="7104
 \mathchardef\PLambda="7103
 \mathchardef\PSigma="7106
 \mathchardef\POmega="710A
 \mathchardef\PUpsilon="7107
                  
 \def\PB      {\ensuremath{B}\xspace}                 
                  
 \def\PD      {\ensuremath{D}\xspace}

 \def\PJ      {\ensuremath{J}\xspace}                 
 \def\PK      {\ensuremath{K}\xspace}

 \def\Pb      {\ensuremath{b}\xspace}                 
 \def\Pc      {\ensuremath{c}\xspace}

 \def\Pi      {\ensuremath{i}\xspace}

}

\makeatletter
\ifcase \@ptsize \relax
  \newcommand{\miniscule}{\@setfontsize\miniscule{4}{5}}
\or
  \newcommand{\miniscule}{\@setfontsize\miniscule{5}{6}}
\or
  \newcommand{\miniscule}{\@setfontsize\miniscule{5}{6}}
\fi
\makeatother

\DeclareRobustCommand{\optbar}[1]{\shortstack{{\miniscule (\rule[.5ex]{1.25em}{.18mm})}
  \\ [-.7ex] $#1$}}

\def\mup        {{\ensuremath{\Pmu^+}}\xspace}
\def\mun        {{\ensuremath{\Pmu^-}}\xspace} 

\def\cquark    {{\ensuremath{\Pc}}\xspace}

\def\bquark    {{\ensuremath{\Pb}}\xspace}

\def\pion   {{\ensuremath{\Ppi}}\xspace}

\def\pip    {{\ensuremath{\pion^+}}\xspace}
\def\pim    {{\ensuremath{\pion^-}}\xspace}
\def\pipm   {{\ensuremath{\pion^\pm}}\xspace}
\def\pimp   {{\ensuremath{\pion^\mp}}\xspace}

\def\kaon    {{\ensuremath{\PK}}\xspace}
  \def\Kbar    {{\kern 0.2em\overline{\kern -0.2em \PK}{}}\xspace}

\def\KorKbar    {\kern 0.18em\optbar{\kern -0.18em K}{}\xspace}
\def\Kz      {{\ensuremath{\kaon^0}}\xspace}
\def\Kzb     {{\ensuremath{\Kbar{}^0}}\xspace}
\def\Kp      {{\ensuremath{\kaon^+}}\xspace}
\def\Km      {{\ensuremath{\kaon^-}}\xspace}
\def\Kpm     {{\ensuremath{\kaon^\pm}}\xspace}

\def\Kstarz  {{\ensuremath{\kaon^{*0}}}\xspace}

\def\Kstar   {{\ensuremath{\kaon^*}}\xspace}

  \def\Dbar    {{\kern 0.2em\overline{\kern -0.2em \PD}{}}\xspace}
\def\D       {{\ensuremath{\PD}}\xspace}

\def\DorDbar    {\kern 0.18em\optbar{\kern -0.18em D}{}\xspace}
\def\Dz      {{\ensuremath{\D^0}}\xspace}

\def\Dstarp  {{\ensuremath{\D^{*+}}}\xspace}

\def\B       {{\ensuremath{\PB}}\xspace}
\def\Bbar    {{\ensuremath{\kern 0.18em\overline{\kern -0.18em \PB}{}}}\xspace}

\def\BorBbar    {\kern 0.18em\optbar{\kern -0.18em B}{}\xspace}

\def\Bu      {{\ensuremath{\B^+}}\xspace}
\def\Bub     {{\ensuremath{\B^-}}\xspace}
\def\Bp      {{\ensuremath{\Bu}}\xspace}
\def\Bm      {{\ensuremath{\Bub}}\xspace}
\def\Bpm     {{\ensuremath{\B^\pm}}\xspace}

\def\jpsi     {{\ensuremath{{\PJ\mskip -3mu/\mskip -2mu\Ppsi\mskip 2mu}}}\xspace}

  \def\Y#1S{\ensuremath{\PUpsilon{(#1S)}}\xspace}

\def\Lbar        {{\ensuremath{\kern 0.1em\overline{\kern -0.1em\PLambda}}}\xspace}
\def\LorLbar    {\kern 0.18em\optbar{\kern -0.18em \PLambda}{}\xspace}

\def\to                 {\ensuremath{\rightarrow}\xspace}

\def\CP                {{\ensuremath{C\!P}}\xspace}
\def\CPT               {{\ensuremath{C\!PT}}\xspace}

\def\AT#1     {\ensuremath{A_{\mathrm{T}}^{#1}}\xspace}           

\def\C#1      {\ensuremath{\mathcal{C}_{#1}}\xspace}                       
\def\Cp#1     {\ensuremath{\mathcal{C}_{#1}^{'}}\xspace}                    
\def\Ceff#1   {\ensuremath{\mathcal{C}_{#1}^{\mathrm{(eff)}}}\xspace}        
\def\Cpeff#1  {\ensuremath{\mathcal{C}_{#1}^{'\mathrm{(eff)}}}\xspace}       
\def\Ope#1    {\ensuremath{\mathcal{O}_{#1}}\xspace}                       
\def\Opep#1   {\ensuremath{\mathcal{O}_{#1}^{'}}\xspace}                    

\newcommand{\tev}{\ifthenelse{\boolean{inbibliography}}{\ensuremath{~T\kern -0.05em eV}\xspace}{\ensuremath{\mathrm{\,Te\kern -0.1em V}}}\xspace}
\newcommand{\gev}{\ensuremath{\mathrm{\,Ge\kern -0.1em V}}\xspace}
\newcommand{\mev}{\ensuremath{\mathrm{\,Me\kern -0.1em V}}\xspace}
\newcommand{\kev}{\ensuremath{\mathrm{\,ke\kern -0.1em V}}\xspace}
\newcommand{\ev}{\ensuremath{\mathrm{\,e\kern -0.1em V}}\xspace}
\newcommand{\gevc}{\ensuremath{{\mathrm{\,Ge\kern -0.1em V\!/}c}}\xspace}
\newcommand{\mevc}{\ensuremath{{\mathrm{\,Me\kern -0.1em V\!/}c}}\xspace}
\newcommand{\gevcc}{\ensuremath{{\mathrm{\,Ge\kern -0.1em V\!/}c^2}}\xspace}
\newcommand{\gevgevcccc}{\ensuremath{{\mathrm{\,Ge\kern -0.1em V^2\!/}c^4}}\xspace}
\newcommand{\mevcc}{\ensuremath{{\mathrm{\,Me\kern -0.1em V\!/}c^2}}\xspace}

\def\invfb   {\ensuremath{\mbox{\,fb}^{-1}}\xspace}

\def\gsim{{~\raise.15em\hbox{$>$}\kern-.85em
          \lower.35em\hbox{$\sim$}~}\xspace}
\def\lsim{{~\raise.15em\hbox{$<$}\kern-.85em
          \lower.35em\hbox{$\sim$}~}\xspace}

\def\sPlot{\mbox{\em sPlot}}

\def\ptot       {\mbox{$p$}\xspace}
\def\pt         {\mbox{$p_{\rm T}$}\xspace}

\def\evtgen     {\mbox{\textsc{EvtGen}}\xspace}

\def\geant      {\mbox{\textsc{Geant4}}\xspace}

\def\photos     {\mbox{\textsc{Photos}}\xspace}

\def\pythia     {\mbox{\textsc{Pythia}}\xspace}

\def\tell1  {TELL1\xspace}
\def\ukl1   {UKL1\xspace}

\newcommand{\eg}{\mbox{\itshape e.g.}\xspace}
\newcommand{\ie}{\mbox{\itshape i.e.}\xspace}

%% file: bhhh-symbols-def.tex
\def\hhh {\ensuremath{{\Bpm \to h^{\pm} h^+ h^-}}\xspace}

\def\pipipi {\ensuremath{{\Bpm \to \pipm \pip \pim}}\xspace}
\def\kpipi {\ensuremath{{\Bpm \to \Kpm \pip \pim}}\xspace}
\def\kkpi {\ensuremath{{\Bpm \to \pipm \Kp \Km }}\xspace}
\def\kkk {\ensuremath{{\Bpm \to \Kpm \Kp \Km}}\xspace}

\def\jpsik {\ensuremath{{\Bpm \to \jpsi \Kpm}}\xspace}
\def\jpsimumu {\ensuremath{{\jpsi \to \mup\mun}}\xspace}

\def\mmkpi {\ensuremath{m^2(\Kpm \pimp)}\xspace}                    
\def\mpipi {\ensuremath{m(\pip \pim)}\xspace}                        
\def\mmpipi {\ensuremath{m^2(\pip \pim)}\xspace}                        
\def\mmkk {\ensuremath{m^2(\Kp \Km)}\xspace}                            
\def\mkk {\ensuremath{m(\Kp \Km)}\xspace}                            
\def\mkpi {\ensuremath{m(\Kpm \pimp)}\xspace}                            
\def\mmkklow {\ensuremath{m^2(\Kp \Km)_{\rm low}}\xspace}     
\def\mkklow {\ensuremath{m(\Kp \Km)_{\rm low}}\xspace}     
\def\mmpipilow {\ensuremath{m^2(\pip \pim)_{\rm low}}\xspace} 
\def\mpipilow {\ensuremath{m(\pip \pim)_{\rm low}}\xspace} 

\def\acp {\ensuremath{A_{\CP}}\xspace}

\def\acpraw {\ensuremath{A_{\rm raw}}\xspace} 
\def\acpn {\ensuremath{A_{\rm raw}^{N}}\xspace}

\def\aprod {\ensuremath{A_{\rm P}}\xspace}
\def\adet {\ensuremath{A_{\rm D}}\xspace}
\def\adetk {\ensuremath{A_{\rm D}^{K}}\xspace}
\def\adetpi {\ensuremath{A_{\rm D}^{\pi}}\xspace}
